\documentclass[reprint,amssymb, amsmath,twocolumn, aps, superscriptaddress, showpacs, footinbib, prb]{revtex4-1}
\usepackage{graphicx}
\usepackage{color}
\usepackage[colorlinks,breaklinks,bookmarks=true,citecolor=blue,linkcolor=red,urlcolor=blue]{hyperref}

\newcommand{\eff}{\text{eff}}
\newcommand{\AFM}{\text{AFM}}
\newcommand{\FM}{\text{FM}}
\newcommand{\jj}{$J_1$--$J_2$}

\newcommand{\degr}{$^\circ$}

\begin{document}

\title{Frustrated spin chain physics near the Majumdar-Ghosh point in szenicsite Cu$_3$(MoO$_4$)(OH)$_4$}

\author{Stefan Lebernegg}
\email{st.lebernegg@gmail.com}
\affiliation{Leibniz Institute for Solid State and Materials Research, 01171 Dresden, Germany}
\affiliation{Max Planck Institute for Chemical Physics of Solids, N\"{o}thnitzer Str. 40, 01187 Dresden, Germany}

\author{Oleg Janson}
\affiliation{Institute of Solid State Physics, Technical University Vienna, Wiedner Hauptstr. 8--10/138, 1040 Vienna, Austria}

\author{Ioannis~Rousochatzakis}
\affiliation{School of Physics and Astronomy, University of Minnesota, Minneapolis, MN 55455, USA}

\author{Satoshi Nishimoto}
\affiliation{Institute for Theoretical Physics, Technical University Dresden, Helmholtzstr. 10, 01069 Dresden, Germany}
\affiliation{Leibniz Institute for Solid State and Materials Research, 01171 Dresden, Germany}

\author{Helge Rosner}
\affiliation{Max Planck Institute for Chemical Physics of Solids, N\"{o}thnitzer Str. 40, 01187 Dresden, Germany}

\author{Alexander A. Tsirlin}
\email{altsirlin@gmail.com}
\affiliation{Experimental Physics VI, Center for Electronic Correlations and Magnetism, Institute of Physics, University of Augsburg, 86135 Augsburg, Germany}

\date{\today}

\begin{abstract}
In this joint experimental and theoretical work magnetic properties of the Cu$^{2+}$ mineral szenicsite Cu$_3$(MoO$_4$)(OH)$_4$ are investigated. This compound features isolated triple chains in its crystal structure, where the central chain involves an edge-sharing geometry of the CuO$_4$ plaquettes, while the two side chains feature a corner-sharing zig-zag geometry. The magnetism of the side chains can be described in terms of antiferromagnetic dimers with a coupling larger than 200\,K. The central chain was found to be a realization of the frustrated antiferromagnetic \jj~chain model with $J_1\simeq 68$\,K and a sizable second-neighbor coupling $J_2$. The central and side chains are nearly decoupled owing to interchain frustration. Therefore, the low-temperature behavior of szenicsite should be entirely determined by the physics of the central frustrated \jj~chain. Our heat-capacity measurements reveal an accumulation of entropy at low temperatures and suggest a proximity of the system to the Majumdar-Ghosh point of the antiferromagnetic \jj~spin chain, $J_2/J_1=0.5$.
\end{abstract}

\maketitle

\section{Introduction}
\begin{figure}[tbp] \includegraphics[width=0.9\linewidth]{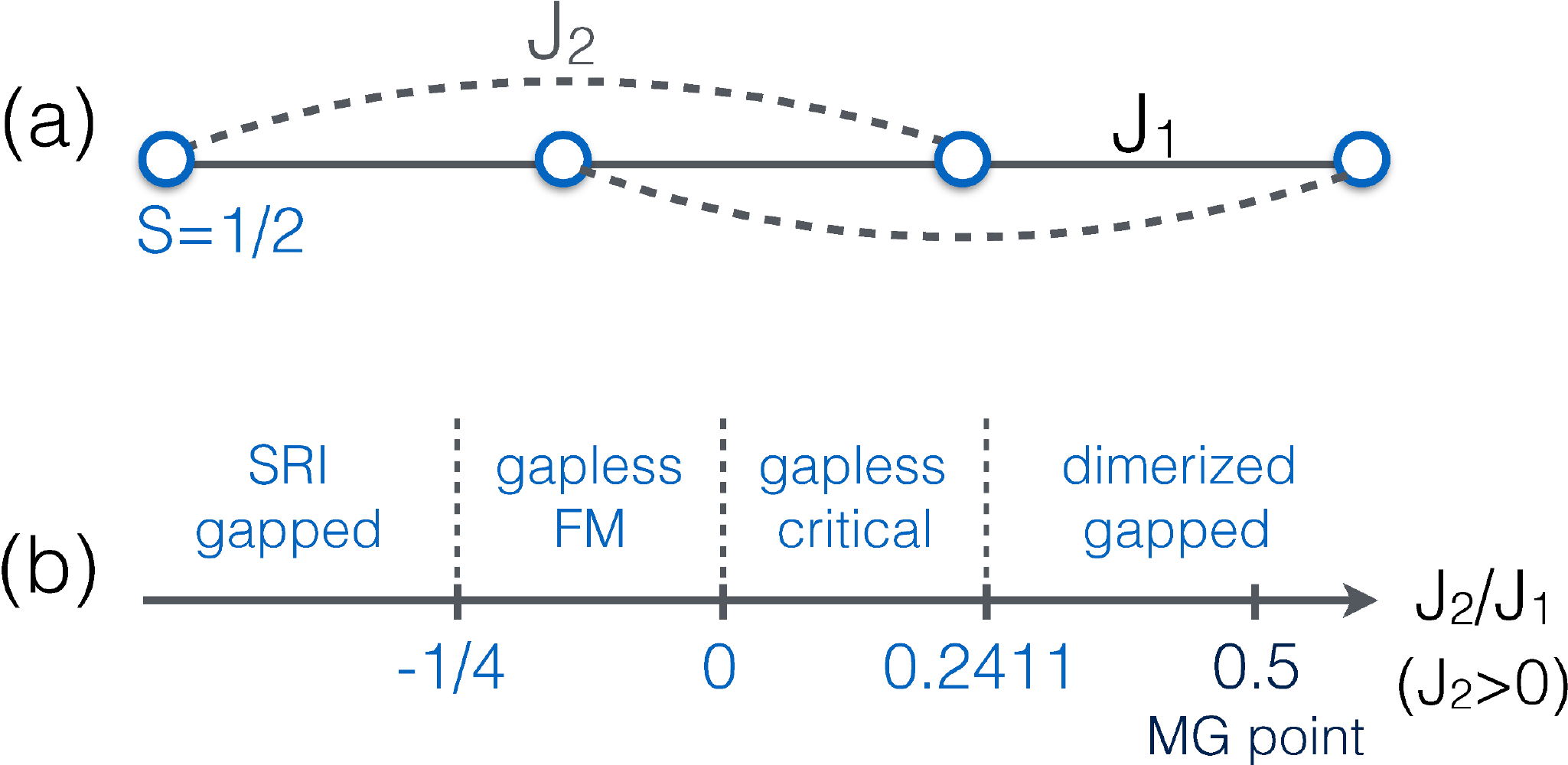}
\caption{\label{j1j2model}(Color online) 
Panel a: the \jj~model. Panel b: magnetic ground state depending on the $J_2/J_1$ ratio ($J_2>0$). SRI denotes a short-range incommensurate phase, MG abbreviates Majumdar-Ghosh and FM means ferromagnetic. $J_1$ and $J_2$ are the nearest-neighbor and next-nearest neighbor exchange couplings, respectively.}
\end{figure}

The study of low-dimensional, particularly one-dimensional spin-1/2 magnets, developed into a field of its own, because these materials provide a unique possibility to study ground and excited states of a large variety of quantum models with very high accuracy~\cite{1d_mag,quant_magn}. Despite their conceptual simplicity, these models provide rich magnetic phase diagrams~\cite{chitra1995,sudan2009} arising from a complex interplay of quantum fluctuations, exchange interactions and lattice topology, which often result in magnetic frustration. In one dimension, the simplest frustrated geometry is the \jj~chain, where $J_1$ and $J_2$ stand for nearest-neighbor (NN) and next-nearest neighbor (NNN) exchange couplings, respectively (Fig.~\ref{j1j2model}). As soon as $J_2$ is antiferromagnetic (AFM), the chain is frustrated, irrespective of the sign of $J_1$. The magnetic ground state (GS) of the system is governed by the frustration ratio $\alpha=J_2/J_1$. 

Real materials entail non-zero interchain couplings that may trigger long-range magnetic order at low temperatures. If $J_1$ is ferromagnetic (FM), i.e., $J_1<0$, ferromagnetic order along the chain is stabilized for the regime $-0.25<\alpha\le0$. At $\alpha=-0.25$, the system undergoes quantum phase transition to an incommensurate spiral state (Fig.~\ref{j1j2model})~\cite{zinke2009,furukawa2010,sirker2011,balents2016}. The resulting spin helix interpolates between the FM chain and the limit of two decoupled AFM chains at $\alpha=-\infty$~\cite{banks2009,lee2012,willenberg2012,pregelj2015}. Materials with the helical order were recently recognized as potential multiferroics~\cite{park2007,naito2007,schrettle2008,seki2010,yasui2011,zhao2012}. In a magnetic field, intricate phase relations have been predicted for the \jj~model~\cite{j1j2_H1,j1j2_H2,sudan2009,zhitomirsky2010,sato2013} and were also studied experimentally~\cite{willenberg2016,weickert2016}. 

In the second type of the frustrated \jj~chains, both $J_1$ and $J_2$ couplings are AFM. This regime is very different from the ferromagnetic one. For $\alpha>\alpha_c\simeq 0.241$~\cite{eggert1996}, a spin gap is opened~\cite{white1996}, and the chain is effectively in the dimerized state. At the Majumdar-Ghosh point ($\alpha=0.5$), an exact ground state is represented by a superposition of spin singlets~\cite{majumdar1969,*majumdar1969b}. The gap opening can be assisted by a structural dimerization. In the spin-Peierls compound CuGeO$_3$ ($\alpha\simeq 0.35$), spin-lattice coupling controls the opening of the spin gap and the formation of soliton phases in external magnetic field~\cite{hase1993,kiryukhin1996}. Interestingly, no gapped AFM $J_1-J_2$ chain material without the spin-lattice coupling has been reported. While the $\alpha>\alpha_c$ regime has been proposed for (N$_2$H$_5$)CuCl$_3$~\cite{brown1979,hagiwara2001,maeshima2003} and KTi(SO$_4$)$_2\cdot$2H$_2$O~\cite{nilsen2008,kasinathan2013}, the presence of a spin gap in these materials remains to be probed experimentally.

In the present work, we demonstrate that the physics of AFM \jj-chains can be observed in the Cu$^{2+}$ mineral szenicsite whose crystal structure features isolated slabs of triple-chains of Cu$^{2+}$ ions. Recent studies of this compound~\cite{szen_1,nmr_szen} suggested a magnetic model of uniform AFM spin chains that, however, was derived empirically and turns out to be insufficient for describing the experimental data. In our study, a consistent microscopic magnetic model is derived from density-functional theory (DFT) calculations. This model fully captures the results of our thermodynamical measurements. Based on the DFT-results, we develop an effective microscopic magnetic model, demonstrating that at low temperatures szenicsite can be described in terms of an AFM \jj~chain with a weak alternation of NNN couplings. Computed exchange couplings as well as our model analysis of the experimental data reveal the $J_2/J_1$ ratio of 0.4--0.5 and place the system in the dimerized phase of the \jj~chain model in close proximity to the Majumdar-Ghosh point. 

The paper is organized as follows: The applied experimental and theoretical methods are described in Sec.~\ref{sec:methods}. Sec.~\ref{sec:structure} contains details of the crystal structure of szenicsite. Exchange couplings calculated within DFT and the effective magnetic model will be presented in Sec.~\ref{sec:magexc}. Sec.~\ref{sec:exp} contains experimental results and their model analysis. Detailed discussion and summary are given in Secs.~\ref{sec:discussion} and~\ref{sec:summary}, respectively. We also compare szenicsite to the isostructural mineral antlerite, Cu$_3$(SO$_4$)(OH)$_4$, that demonstrates idle-spin behavior~\cite{antlerite_2003,antlerite_2007}.



\section{Methods}
\label{sec:methods}
All experimental data were collected on a natural polycrystalline sample (Fig.~\ref{struct}) of szenicsite from the type locality of this mineral, the Jardinera No.1 mine, Atacama region in Chile. Particularly for natural compounds, sample quality is a crucial issue, which we thoroughly checked with laboratory powder x-ray diffraction (XRD) (Huber G670 Guinier camera, CuK$_{\alpha\,1}$ radiation, ImagePlate detector, $2\theta\,=\,3-100^{\circ}$ angle range). This analysis yielded pure szenicsite without any detectable other phases~\cite{supplement}. Additionally, high-resolution XRD data were collected at room temperature and at 80\,K at the ID31 beamline of the European Synchrotron Radiation Facility (ESRF, Grenoble) at a wavelength of about 0.4\,\r{A}. No detectable structural changes were observed, and the refined structural parameters agreed nicely with the available single crystal data~\cite{xxstr}.

Magnetization was measured using Quantum Design (QD) SQUID MPMS in static fields up to 5\,T and using vibrating sample magnetometer (VSM) setup of QD PPMS up to 14\,T in the temperature range of 1.8--400\,K. Measurements up to 50\,T were performed in pulsed fields at the high magnetic field laboratory of the Helmholtz-Zentrum Dresden-Rossendorf (HZDR).
Heat capacity data were acquired with the QD PPMS in fields up to 14\,T and at temperatures down to 0.5\,K using the $^3$He insert.

Electronic structure calculations within DFT were performed with the full-potential local-orbital code \texttt{fplo9.07-41}~\cite{fplo} in combination with the local density approximation (LDA)~\cite{pw92}, generalized gradient approximation (GGA)~\cite{pbe96}, and local spin density approximation (LSDA)+$U$ methods~\cite{lsdau1,lsdau2}. A well-converged $k$-mesh of 4$\times$6$\times$12 points was used for LDA and GGA runs, while supercells were computed for about 20 $k$-points in the irreducible wedge of the first Brillouin zone.

Accurate hydrogen positions, which are essential for calculating exchange couplings~\cite{malachite,clinoclase}, are hard to obtain experimentally with XRD. The positions provided in the literature~\cite{xxstr} were determined using empirical constraints. We thus determined the atomic H-positions by a structure optimization with respect to the total energy using the GGA DFT-functional. Thereby, lattice parameters and the positions of all other atoms in the unit cell are fixed to their experimental values. 

Exchange coupling constants were calculated within DFT following two different, complementary strategies. First, an LDA band structure is calculated. Since strong electronic correlations are not properly described in LDA, it typically yields spurious metallic ground state. The half-filled bands at the Fermi level, however, allow identifying crucial exchange pathways and are sufficient for the calculation of the low-energy part of the magnetic excitation spectrum. A projection of these bands onto a tight-binding (TB) model yields transfer integrals $t_{ij}$, which we calculated in the present case as off-diagonal Hamiltonian matrix-elements between Cu-centered Wannier functions (WFs). Next, to account for the strong electron correlations, the TB-model is supplemented by the effective on-site Coulomb repulsion, $U_{\text{eff}}$, of electrons in the Wannier orbitals. This yields the single-band Hubbard model: 
$\hat{H}=\hat{H}_{\rm TB}+U_{\text{eff}}\sum_{i}\hat{n}_{i\uparrow}\hat{n}_{i\downarrow}$. Subsequently, the Hubbard model is mapped onto a Heisenberg model 
\begin{equation} 
\hat{H}=\sum_{\left\langle ij\right\rangle}J_{ij}\hat{S_{i}}\cdot\hat{S_{j}},
\end{equation}
where the summation is over lattice bonds $\left\langle ij\right\rangle$. This procedure is justified at $t_{ij}\gg U_{\text{eff}}$ and half-filling, as appropriate for szenicsite (Table~\ref{tJ}). AFM contributions to the exchange constants $J_{ij}$ are then obtained in second order as $J_{ij}^{\AFM}=4t_{ij}^2/U_{\eff}$, where we used $U_{\eff}=4.5$\,eV according to our previous studies of cuprates~\cite{diaboleite,malachite}.

Alternatively, the full exchange constants, $J_{ij}=J_{ij}^{\AFM}+J_{ij}^{\FM}$, containing both FM and AFM contributions, are obtained by including strong electronic correlations in the self-consistent procedure. Within DFT, the $J_{ij}$ are calculated from differences between total energies of various collinear (broken-symmetry) spin states~\cite{bs2}, which are evaluated in spin-polarized supercell calculations using the LSDA+$U$ formalism. The double counting is corrected with an "around-mean-field" (AMF) approximation as implemented in \texttt{fplo9.07-41}. The on-site Coulomb repulsion of the atomic Cu(3$d$) orbitals, $U_d$, was set to $6.5\pm0.5$\,eV and the on-site Hund's exchange, $J_d$, was fixed to 1.0\,eV, the parameter set that is typically used for calculations of cuprates with \texttt{fplo9.07-41}~\cite{dioptase,clinoclase,callaghanite}.

Exact diagonalization (ED) and density-matrix renormalization group (DMRG) simulations of one-dimensional spin models were performed using the code \textsc{alps-1.3}~\cite{ALPS}. Periodic and open boundary conditions were used for ED and DMRG, respectively.

To analyze the specific heat measurements we applied the transfer-matrix renormalization group (TMRG) technique.~\cite{wang1997,shibata1997} In our calculations, about 200 states were retained in the renormalization procedure and the truncation error was less than $10^{-4}$ down to $T=0.01|J1|$.

\section{Crystal structure}
\label{sec:structure}
\begin{figure*}[tbp] \includegraphics[width=17.1cm]{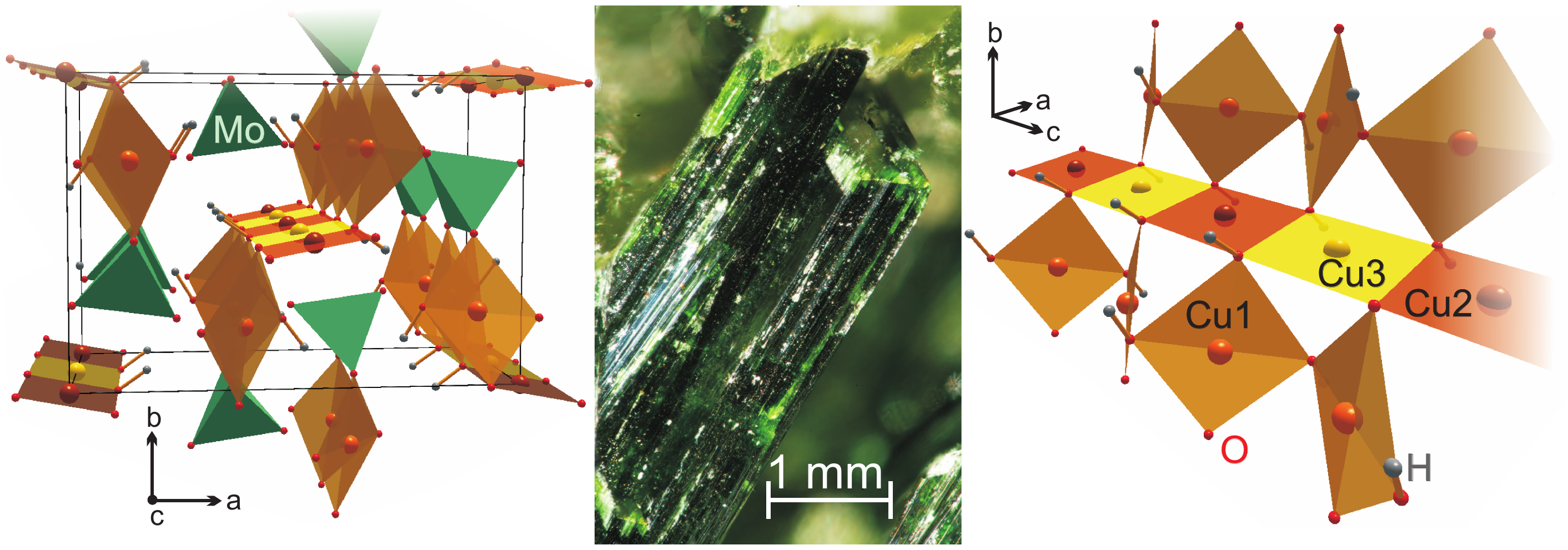}
\caption{\label{struct}(Color online) 
The crystal structure of szenicsite, Cu$_3$(MoO$_4$)(OH)$_4$, with the CuO$_4$ plaquettes of the three different Cu-positions shown in orange, red and yellow. The MoO$_4$ tetrahedra are displayed in green. The central picture shows a natural, dark-green crystal of szenicsite from Jardinera No.1 mine, Atacama region, Chile.}
\end{figure*}

The extremely rare, dark-green Cu-mineral szenicsite, discovered in 1993, crystallizes in the orthorhombic space group $Pnnm$ with the lattice parameters $a=12.559$\,\r{A}, $b=8.518$\,\r{A} and $c=6.072$\,\r{A}~\cite{xxstr}. Its unit cell contains 3 different Cu-positions. The CuO$_4$ plaquettes of the Cu2 and Cu3 sites share common edges and form planar chains running parallel to the $c$-direction (Fig.~\ref{struct}). These chains are decorated by two strongly buckled side-chains consisting of corner-sharing plaquettes formed by Cu1 atoms. The resulting triple-chains are the central building block of szenicsite. Relevant Cu--O--Cu bridging angles and Cu--Cu distances, both of crucial importance for the exchange couplings, are given in Table~\ref{tJ}. The MoO$_4$ tetrahedra bond to the outer edges of the triple chains, but do not bridge different triple-chains with each other. These chains thus remain isolated. 

The hydrogen positions, particularly the O--H distance and the angle between the O--H bond and the CuO$_4$ plaquette plane, are essential for the exchange couplings~\cite{clinoclase,malachite,callaghanite}. For example, a large out-of-plane angle of the hydrogen attached to the bridging oxygen reduces the transfer integral between the two Cu-sites and, accordingly, reduces $J^{\text{\AFM}}_{ij}$, thus favoring a FM coupling~\cite{clinoclase,callaghanite,ruiz97_2}. The H atomic positions provided in the literature were obtained under empirical constraints~\cite{xxstr} and, therefore, cannot provide the required accuracy. From optimization within DFT using the GGA-functional, we obtained the atomic H-positions listed in Table~\ref{Hpos}. In the case of szenicsite, the tentative H-positions of Ref.~\onlinecite{xxstr} were very close to our optimized ones. However, it is worth noting that even small changes of the positions of H may have crucial effect on the exchange pathways, particularly when FM and AFM contributions of the same magnitude are involved. 

\begin{table}[tbp]
\begin{ruledtabular}
\caption{\label{Hpos} 
Fractional coordinates of hydrogen as obtained from the GGA structure optimization. Lattice parameters and positions of all other atoms were fixed to those of the room-temperature structure determined by single-crystal XRD~\cite{xxstr}.}
\begin{tabular}{c c c c}
atom & $x/a$ & $y/b$ & $z/c$ \\ \hline
H1 & $-0.4895$   &  0.1346   &  0         \\
H2 &  0.2330   &  0.3410   &  $\frac12$       \\ 
H3 &  0.3343   & $-0.4618$   &  0.26042   \\ 
\end{tabular}
\end{ruledtabular}
\end{table}

The crystal structure of szenicsite is closely related to that of antlerite, Cu$_3$(SO$_4$)(OH)$_4$~\cite{xxstr}, containing sulphate instead of molybdate anionic groups. The magnetic structure of antlerite has been intensively investigated~\cite{antlerite_2009,antlerite_2011,antlerite_2012,antlerite_2013} due to a proposed idle spin behavior~\cite{antlerite_2007}. However, despite large structural similarities, the magnetic properties of these two compounds are very different. In Sec.~\ref{sec:discussion}, the origin of this dissimilarity will be analyzed on the basis of structural and chemical differences between the two compounds.

\section{Magnetic exchange couplings}
\label{sec:magexc}

\subsection{\label{tb}LDA results and the tight-binding model}

\begin{figure}[tb] \includegraphics[width=8.6cm]{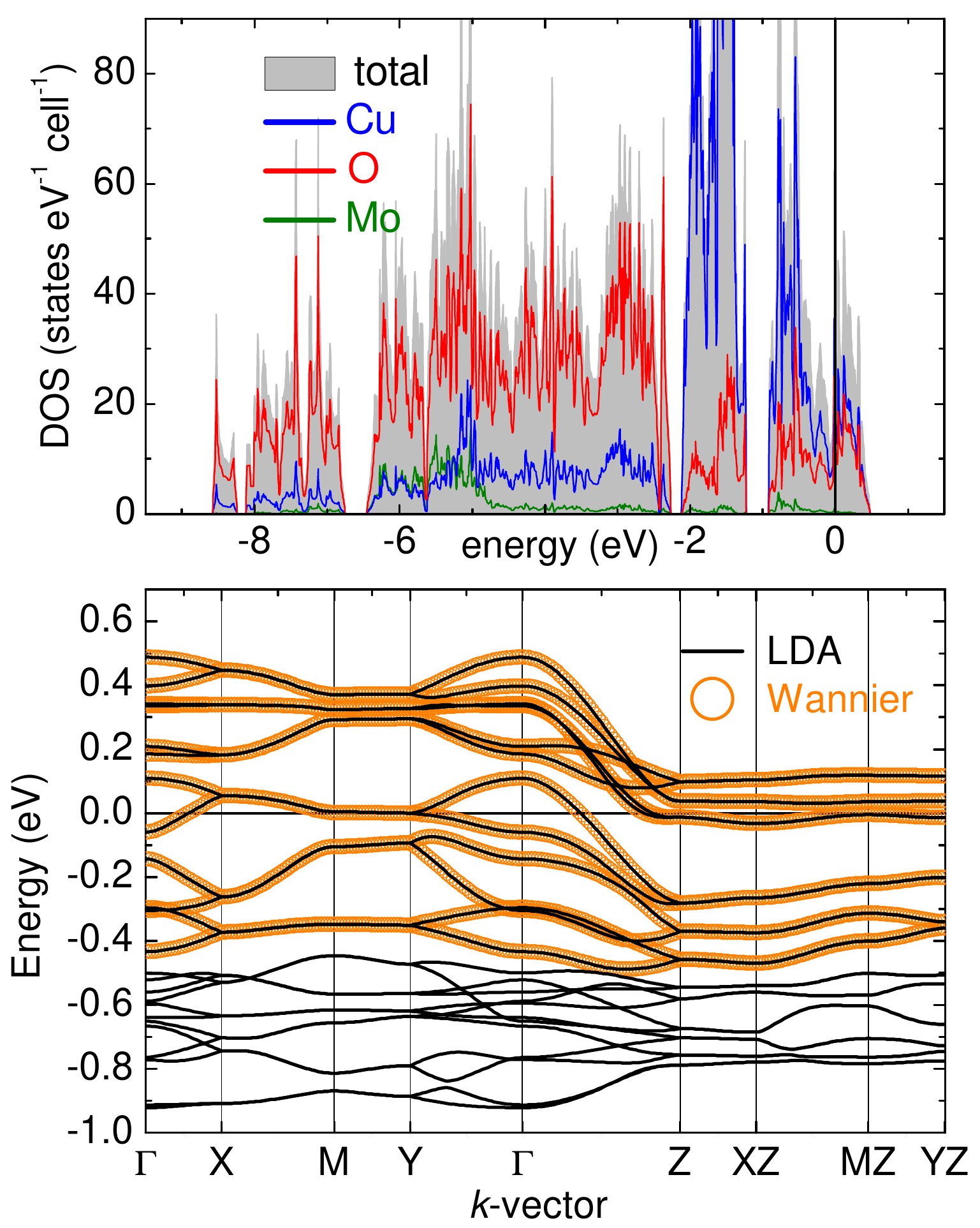}
\caption{\label{lda}(Color online) Results from the LDA calculations. The top panel shows the total and partial density of states (DOS). In the lower panel, the 12 LDA bands around the Fermi level are shown. They are perfectly reproduced with the bands calculated with Cu-centered Wannier functions, denoted as "Wannier".}
\end{figure}

LDA calculations yield a broad valence band complex with the width of about 9\,eV (Fig.~\ref{lda}), which is very typical for cuprates~\cite{clinoclase,callaghanite}. It arises from strong interactions between the O(2$p$) orbitals and the partially occupied Cu(3$d$) shell. Of interest for low-energy magnetic properties are only the partially filled bands around the Fermi level, which have antibonding $pd\sigma$* character. Local coordinate systems on the 12 Cu$^{2+}$-sites per unit cell (the $x$- and $z$-axis are chosen as one of the Cu--O bonds and the direction perpendicular to the CuO$_4$ plaquette plane, respectively) facilitate the analysis of the isolated $pd\sigma$* complex between $-1$ and 0.6\,eV, consisting of 24 bands, with respect to atomic orbital contributions. 

The upper 12 bands are half-filled and predominantly of Cu($3d_{x^2-y^2}$) character, while the lower 12 bands are fully occupied and feature dominant contributions from Cu($3d_{3z^2-r^2}$) orbitals. The separation of these two types of $pd\sigma$* bands arises from the Jahn-Teller distortion of the CuO$_6$ octahedra resulting in a 4+2 coordination with the CuO$_4$ plaquettes spanned by the four short bonds. A projection of the 12 half-filled bands onto local Cu(3$d_{x^2-y^2}$) orbitals yields 12 Cu-centered Wannier functions. A perfect agreement between the bands calculated within LDA and those obtained from the WFs (Fig.~\ref{lda}) is essential for the calculation of reliable transfer integrals $t_{ij}$. It also demonstrates that on the Cu-sites only Cu(3$d_{x^2-y^2}$) orbitals are involved in the magnetism, while the Cu($3d_{z^2-r^2}$)-orbitals are fully filled and remain inactive. 

Table~\ref{tJ} provides all the $t_{ij}$ parameters within the triple chains and the corresponding AFM contributions to the exchange couplings, $J_{ij}^{\text{AFM}}$. Transfer integrals between different triple-chains are below 20\,meV, i.e. $J_{ij}^{\AFM}<5$\,K. 

The calculations reveal a strong NN transfer $t_1$ yielding an AFM coupling of about 150\,K within the central chain, similar to the values found in other compounds with edge-sharing geometry and bridging Cu--O--Cu angles of about 100$^{\circ}$~\cite{clinoclase}. The alternation of the NNN couplings, $J_2$ and $J_2'$, is quite unusual for a \jj-chain but arises naturally from the two crystallographically nonequivalent Cu-sites Cu2 and Cu3. By far less trivial, regarding the similar bridging angles and bonding distances (Table~\ref{tJ}), are the very different strengths of the NN transfers $t_D$ and $t_D'$ within the corner-sharing side chains. These chains communicate with the central chain by sizable $J^{\text{AFM}}_{a,b}$ of 60--90\,K. 

While this LDA-based analysis provides a valuable first idea about the leading exchange pathways and magnetic dimensionality, it lacks the FM contributions which should play an important role, in particular for NN couplings with Cu--O--Cu bridging angles close to $100^{\circ}$~\cite{gka1,*gka2,*gka3}.

\begin{table}[tbp]
\begin{ruledtabular}
\caption{\label{tJ} 
The hopping integrals $t_{ij}$ (in meV) and the AFM contributions to the exchange constants $J^{\text{AFM}}_{ij}=4t_{ij}^2/U_{\text{eff}}$ (in~K), where $U_{\text{eff}}$\,=\,4.5\,eV. The Cu--Cu distances, $d_{\text{Cu--Cu}}$, and the Cu--O--Cu bridging angles are given in \r{A} and deg, respectively. The $J_{ij}$ (in K) are obtained from LSDA+$U$ calculations using $U_d=6.5\pm0.5$\,eV and $J_d=1.0$\,eV. Exchange couplings are shown in Fig.~\ref{hopp}.}
\begin{tabular}{c c c c r r c}
        & Cu-type  &  $d_{\text{Cu--Cu}}$ & Cu--O--Cu      & $t_{ij}$ & $J^{\text{AFM}}_{ij}$ & $J_{ij}$     \\ \hline
$J_1$   & Cu2--Cu3  & 3.036    &  99.3           & $152$    & 237             & $52\pm20$  \\ 
$J_2$   & Cu2--Cu2  & 6.072    &                 & $70$     & 51              & $30\pm3$   \\ 
$J_2'$  & Cu3--Cu3  & 6.072    &                 & $59$     & 36              & $26\pm3$   \\\hline
$J_D'$  & Cu1--Cu1  & 3.029    & 103.4           & $-33$    & 11              & $-15\pm2$  \\ 
$J_D$   & Cu1--Cu1  & 3.043    & 105.2           & $-149$   & 230             & $245\pm30$ \\
$J_{2D}$& Cu1--Cu1  & 6.072    &                 & $53$     & 29              & $5\pm1$    \\\hline  
$J_a$   & Cu1--Cu3  & 3.208    & 103.9           & $92$     & 87              & $22\pm5$   \\ 
$J_b$   & Cu1--Cu2  & 3.212    & 104.3           & $-81$    & 67              & $14\pm2$   \\ 
\end{tabular}
\end{ruledtabular}
\end{table}

\begin{figure}[tb] \includegraphics[width=8.6cm]{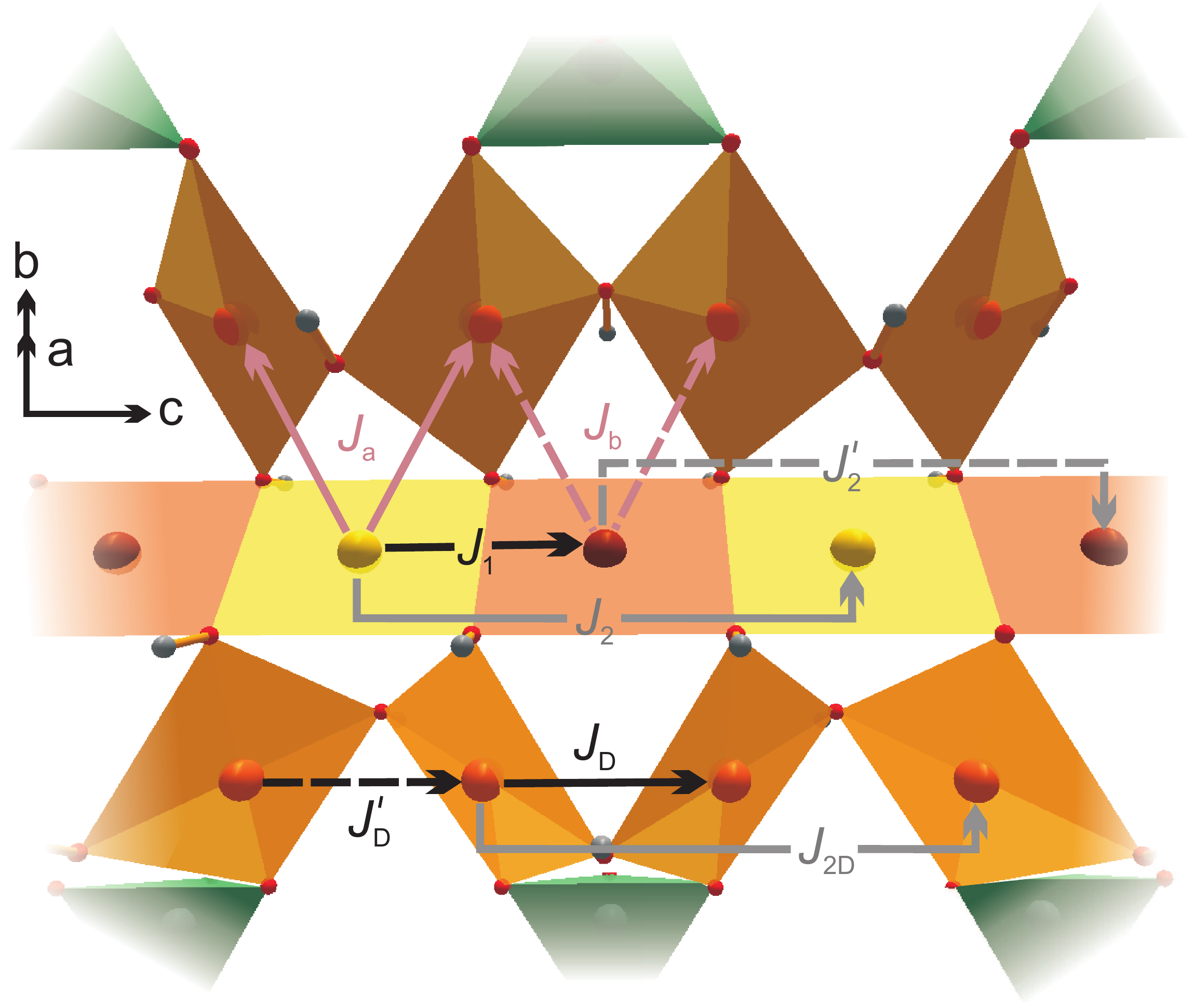}
\caption{\label{hopp}(Color online) Exchange pathways within the structural triple-chain of szenicsite. For an explanation of the color coding, see Fig.~\ref{struct}.}
\end{figure}

\subsection{\label{ldau}LSDA+$U$ results and effective magnetic model}

The exchange couplings computed with LSDA+$U$ comprise both AFM and FM contributions, thus yielding the total $J_{ij}$ given in the last column of Table~\ref{tJ}. As expected with respect to the bridging angles (see Sec.~\ref{sec:discussion}), $J_1$ is considerably reduced by FM contributions resulting in $J_1\simeq 52$\,K and $\alpha \approx 0.5$ for the central chain. By contrast, $J_D$ is barely affected by FM contributions and remains the leading exchange coupling in szenicsite. $J_D'$ is obtained weakly ferromagnetic. The couplings $J_a$ and $J_b$ between the central chain and the side chains are relatively weak and frustrated.

According to the large ratio $J_D/|J_D'|>16$ and the very weak second-neighbor coupling $J_{2D}$, the side chains are reduced to AFM dimers that will form singlet state at low temperatures. Promoting this singlet into a triplet, locally costs a large amount of energy $J_D$, which is barely renormalized by quantum fluctuations driven by, e.g. $J_D'$, which are of the order $(J_D')^2/J_D$.

\begin{figure}[t] \includegraphics[width=0.8\linewidth]{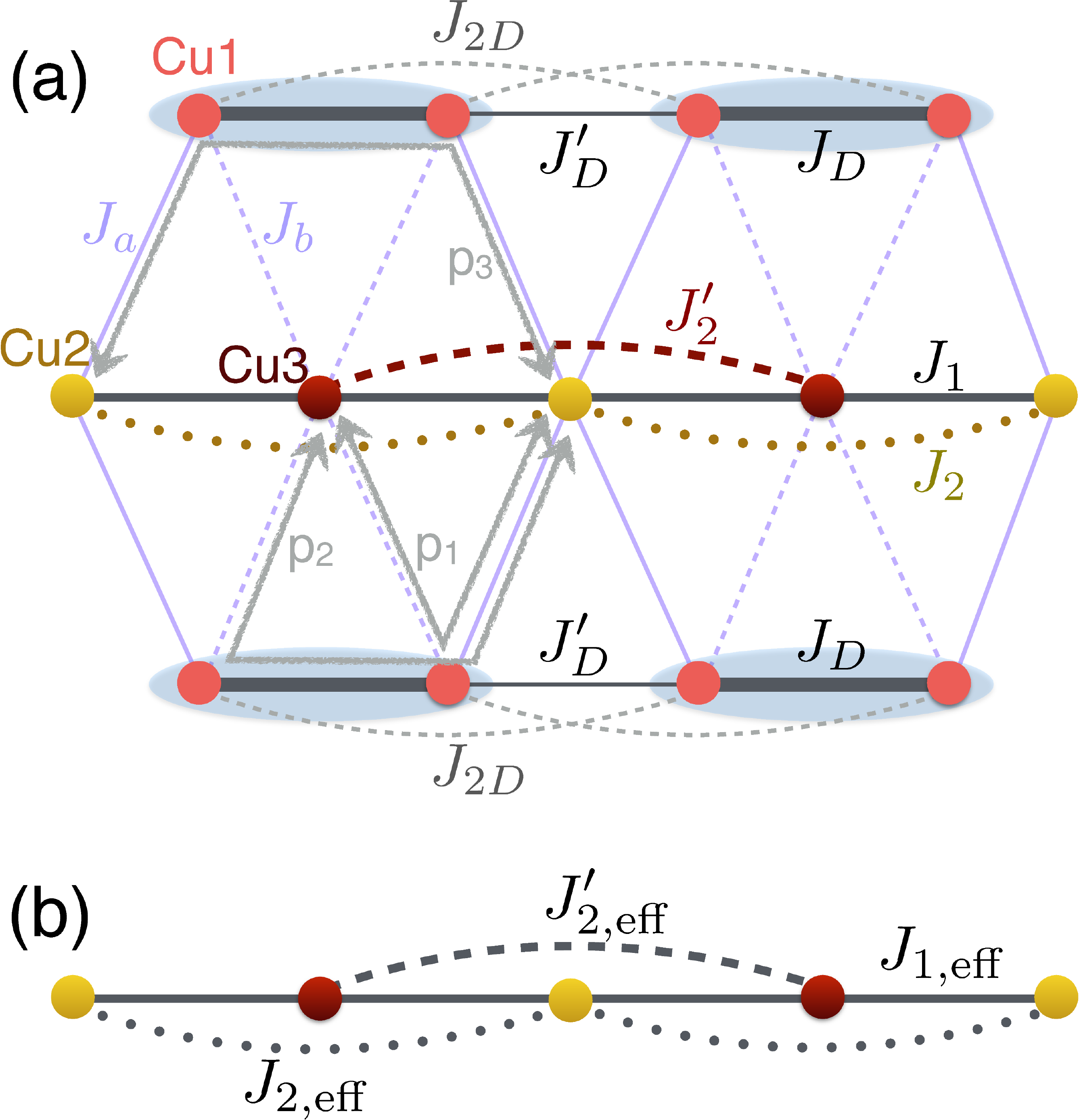}
\caption{\label{EffModel}(Color online) 
Exchange pathways and second-order virtual processes, $p_1-p_3$ (see main text), within the triple-chain of szenicsite are depicted in panel (a). Panel (b) shows the effective spin model of the central Cu2--Cu3 chain.}
\end{figure}

The much smaller energy scale of the exchange couplings involving the central (Cu2--Cu3) chain allows for their treatment independent of the AFM dimers when the temperature is sufficiently low. The respective effective magnetic model for the central chain has been derived in terms of renormalized exchanges $J_{1,\text{eff}}$, $J_{2,\text{eff}}$ and $J_{2,\text{eff}}'$, which are shown in Fig.~\ref{EffModel}. The renormalized interactions are obtained by integrating out the high-energy triplet contributions virtually excited by the interchain couplings $J_a$ and $J_b$. This step is typically performed up to second order of perturbation theory (e.g., Fig.~7 in Ref.~\onlinecite{decoratedSS}). Indeed, the second-order corrections to the NN coupling $J_1$ vanish due to a destructive interference of two types of virtual processes ($p_1$ and $p_2$ in Fig.~\ref{EffModel}~(a)), where $p_1$ and $p_2$ involve one and two sites of the $J_D$ dimer, respectively. For the $J_2$ coupling, we find two virtual paths $p_3$ of the same type, which involve the two Cu1 side chains, leading to a finite correction $J_a^2/J_D$ of only about 2\,K according to the exchange couplings given in Table~\ref{tJ}. Eventually, there are no corrections to $J_2'$ because of the lack of second-order virtual processes. Owing to the minimal renormalzation, we refer only to $J_1$, $J_2$ and $J_2'$ in the further discussion.

Therefore, according to our DFT results and the ensuing effective magnetic model, the magnetic behavior of szenicsite at low temperatures is determined by the central Cu2--Cu3 chains and can be described in terms of an AFM $J_1-J_2-J_2'$ spin model (Fig.~\ref{EffModel}). The Cu1 side chains are reduced to AFM dimers, which should govern the magnetic behavior at higher temperatures, where the Cu2--Cu3 chains are in the Curie-Weiss-like regime.  
    
This result is in striking contrast to the empirically derived model of Ref.~\onlinecite{szen_1}, where purely AFM couplings with $J_D=J_D'=J_1$ were assumed, and a uniform chain model was accepted, even though it could not reproduce the experimental susceptibility data. The LSDA+$U$ results will be challenged in the next section, where our proposed effective microscopic magnetic model serves as basis for the interpretation of thermodynamical data. 

\section{Experimental data and model analysis}
\label{sec:exp}

\subsection{\label{sec:chi} Magnetic susceptibility}

Susceptibility data for szenicsite feature a broad maximum at about 80\,K. The upturn below 30\,K arises from paramagnetic imperfections. No field dependence except for the suppression of the low-T upturn in higher magnetic fields was observed~\cite{supplement}. 

For fitting the susceptibility data we use the general expression $\chi(T)=\chi_0+C_{\text{imp}}/T+\chi_{\text{spin}}(T)$, where $\chi_0$ accounts for the temperature-independent contribution, the second term stands for paramagnetic impurities, and the third term describes the intrinsic spin contributions. Different models for $\chi_{\text{spin}}$ are employed (Table~\ref{sus_fit}), where the number of parameters is limited to three. Models with more parameters produce ambiguous fits and are, thus, discarded. 

\begin{figure}[tb] \includegraphics[width=8.6cm]{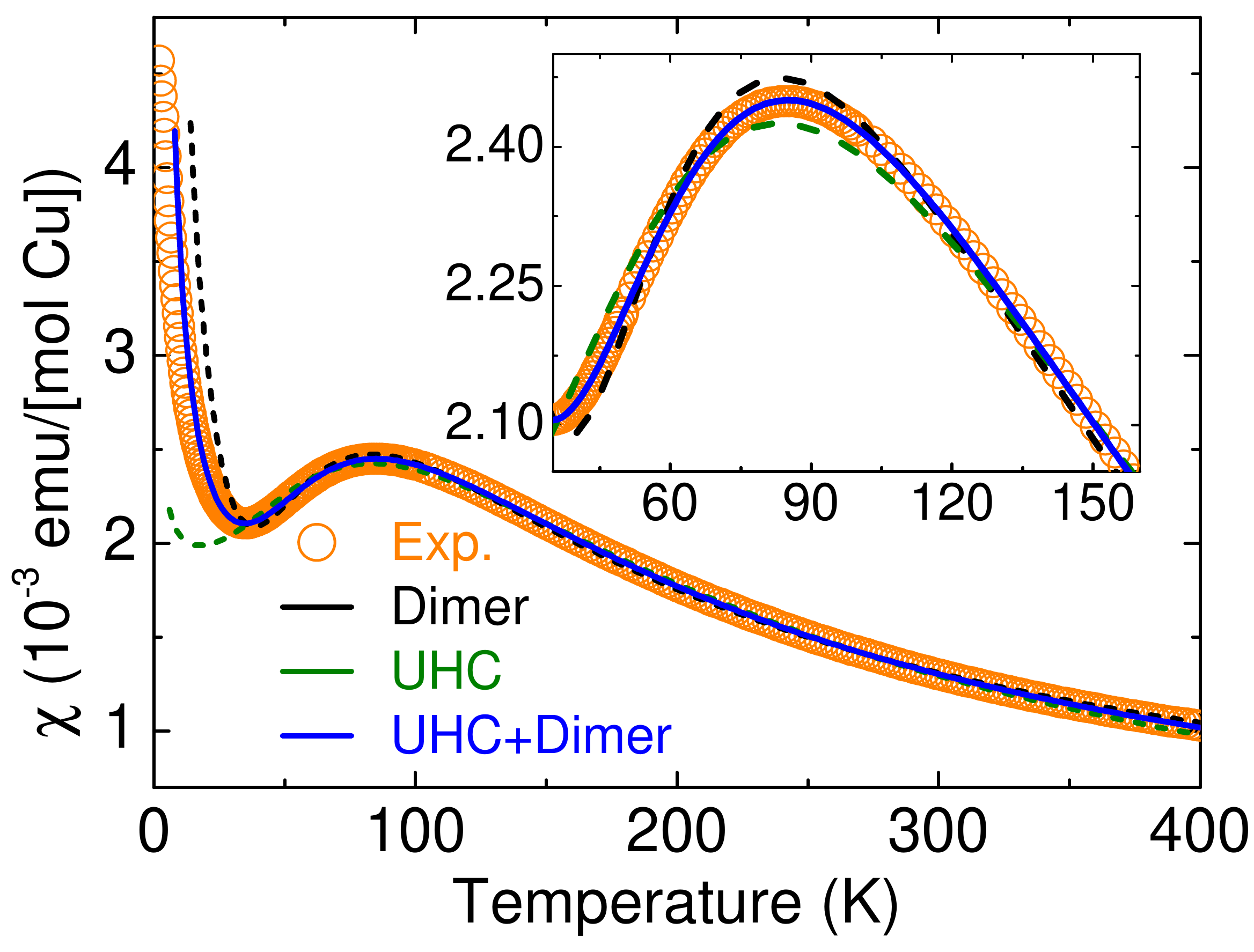}
\caption{\label{susfitF}(Color online) Experimental susceptibility data measured in the applied field of 5\,T. The dashed and solid lines denote the fits with different models (see Table~\ref{sus_fit} and main text).}
\end{figure}

As expected, the least satisfactory result is obtained with the uniform Heisenberg chain (UHC) model proposed in Ref.~\onlinecite{szen_1}, where the $g$-factor is much too large for the Cu$^{2+}$ ion. Moreover, there are deviations between the fitted and experimental curves close to the susceptibility maximum (Fig.~\ref{susfitF}). A simple AFM dimer model provides a better fit, because $J_D$ is the leading energy scale of the system (Table~\ref{tJ}). However, the $g$-value is now below 2.0, and the susceptibility maximum is again not well reproduced. By combining the dimer and UHC models according to $\chi_{\text{spin}}=(2\cdot\chi_{\text{dimer}}+\chi_{\text{UHC}})/3$ (the number of dimers corresponds to their number per unit cell) and using the same $g$-factor for all Cu sites, we arrive at an excellent fit of the experimental curve. In this model, the dimers represent the side chain, and UHC represents the central chain. Despite the simple nature of the model, the estimated exchange couplings $J_D$ and $J_1$ agree quite well with those from LSDA+$U$ calculations (Table~\ref{tJ}), which lends strong support to our proposed microscopic magnetic model. The $g$-factor of 2.18 is within the typical range for cuprates~\cite{malachite,callaghanite}. 
The impurity content of 3.6\% (for spin-1/2 impurities) can be estimated from $C_{\text{imp}}$. 

Although the coupling $J_2$ can be included in the model as well, its value can not be determined unambiguously from the susceptibility fit, because different values of $J_2$, including $J_2=0$, provide an equally good susceptibility fit down to low temperatures.

\begin{table}[tbp]
\begin{ruledtabular}
\caption{\label{sus_fit} 
Parameters obtained by fitting the experimental susceptibility data $\chi(T)$, collected at a magnetic field of 5\,T, with different models (see main text). The "Dimer" and "UHC" denote the dimer and uniform Heisenberg chain models, respectively. $\chi_0$ and $C_{\text{imp}}$ stand for the temperature-independent contribution and paramagnetic impurity contribution, respectively.}
\begin{tabular}{c c c c c c} 
Model     &  $J_D$   & $J_1$   &  $g$  &  $\chi_0$  &  $C_{\text{imp}}$ \\
          &  (K)     & (K)     &       & ($10^{-4}$\,emu/mol) & (emu\,K/mol) \\ \hline
Dimer+UHC &  195.2  &  67.6  &  2.18 &  $-2.77$     &  0.0402   \\
Dimer     &  166.0  &         &  1.96 &   0.99     &  0.0579   \\
UHC       &          & 131.8  &  2.53 &  $-2.82$     &  0.0255   \\        
\end{tabular}
\end{ruledtabular}
\end{table}

\subsection{\label{sec:cp} Specific heat measurements}
\begin{figure}[tb] \includegraphics[width=8.6cm]{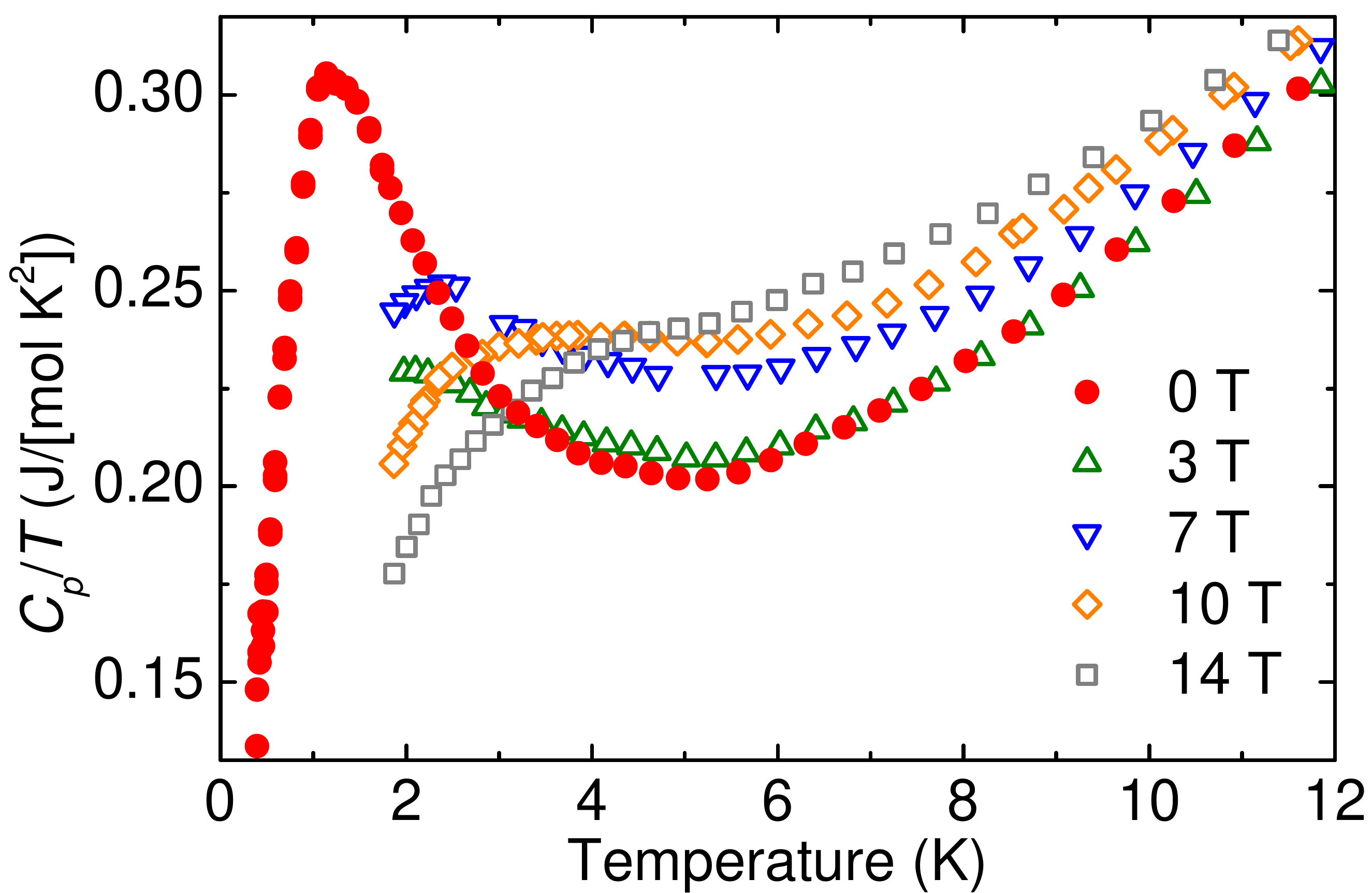}
\caption{\label{cpT}(Color online) Experimental $C_p/T(T)$ data for szenicsite collected in different magnetic fields between 0--14\,T.}
\end{figure}

No signs of anomalies or magnetic phase transitions have been previously observed in the specific heat data for szenicsite measured down to 1.9\,K~\cite{szen_1}. We performed measurements at even lower temperatures, down to 0.5\,K, in order to probe the magnetic entropy of the system. Further measurements in elevated fields up to 14\,T were carried out down to 2\,K. Below 2\,K, a broad shoulder is visible (see supplementary material~\cite{supplement}), which becomes a prominent peak when plotting $C_p/T(T)$. In zero magnetic field, it has a maximum at 1.2\,K (Fig.~\ref{cpT}) and shifts to higher temperatures and broadens with increasing magnetic field. Such a peak is in fact expected for an AFM \jj-chain proximate to the Majumdar-Ghosh point $\alpha=0.5$~\cite{j1j2_AFM_dmrg}. 

\begin{figure}[tb] \includegraphics[width=8.6cm]{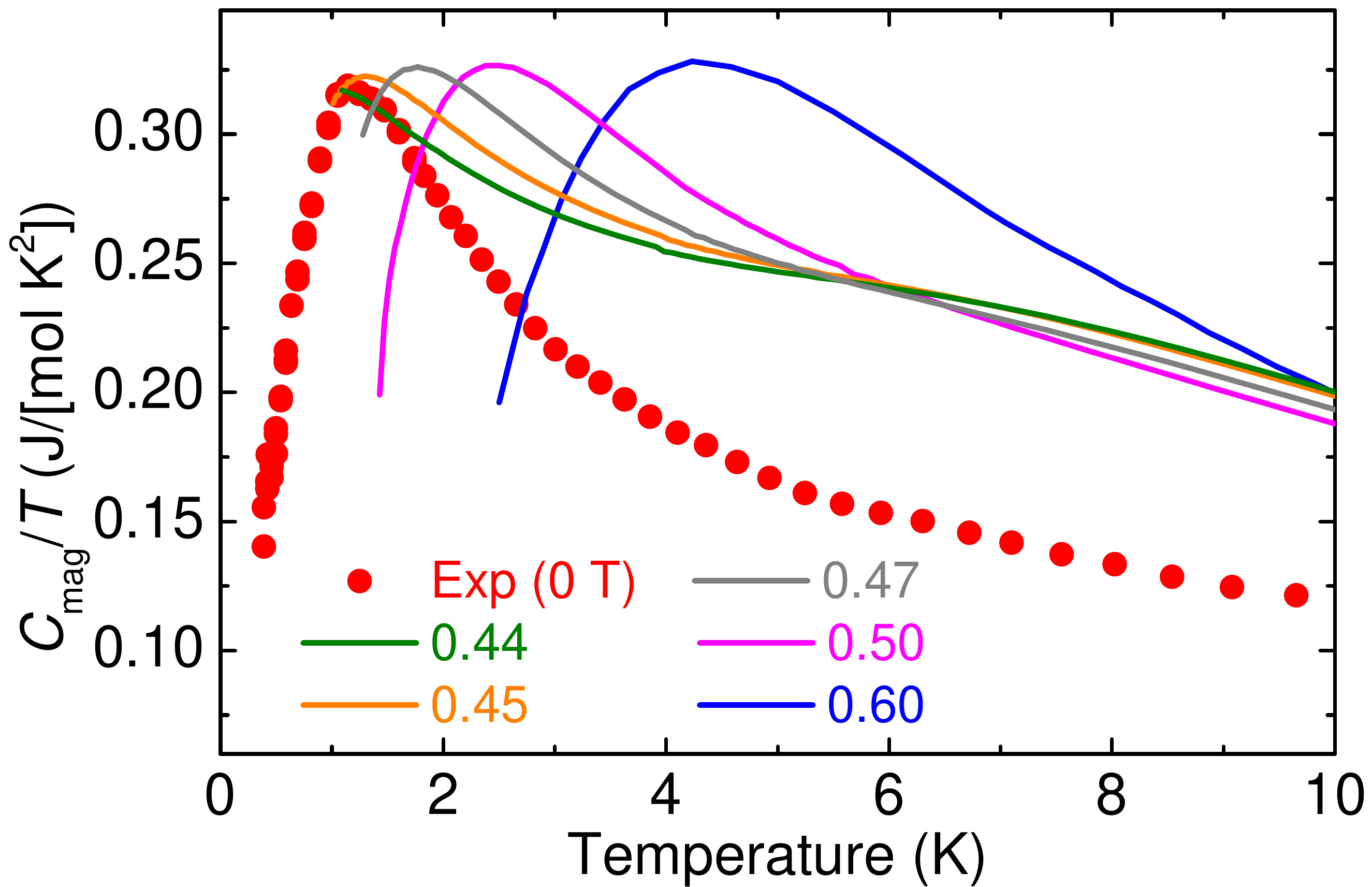}
\caption{\label{cpT_tmrg}(Color online) Comparison of the experimental zero-field $C_{\text{mag}}/T$ data for szenicsite with TMRG simulations of $C/T$ for different frustration ratios $\alpha=0.44$, 0.45, 0.47, 0.5 and 0.6. $C_{mag}$ is the magnetic contribution to the specific heat.}
\end{figure}

According to the data provided in Ref.~\onlinecite{j1j2_AFM_dmrg}, the position of the peak maximum in szenicsite corresponds to the $\alpha$ ratio between 0.4 and 0.5. We performed some more TMRG runs to narrow this range and arrived at a ratio of about 0.45 with $J_1=44$\,K in reasonable agreement with the DFT results. Considerably lower values of $\alpha$ can be excluded, because they result in large deviations in either peak height or peak position. A comparison of the computed data and the experimental magnetic $C_{\text{mag}}/T$ result is shown in Fig.~\ref{cpT_tmrg}. For comparison, we used the magnetic contribution $C_{\text{mag}}$ that was obtained by subtracting the lattice contribution $C_{\text{lat}}$ from the measured $C_p$ data, where $C_{\text{lat}}(T)$ was approximated by fitting a polynomial~\cite{supplement} $C_{\text{lat}}(T)=\sum \limits_{n=3}^{n=7} c_nT^n$, proposed by Johnston \textit{et al.}~\cite{johnston2000}, to the $C_p(T)$ data in the temperature range of 15--39\,K~\cite{supplement}.

The peak position and height is nicely reproduced in the simulation, but the peak shapes do not agree very well. We tested the effect of lowering the $J_2'/J_2$ ratio and found a moderate shift of the peak to lower temperatures. However, for ratios $J_2'/J_2>0.8$ expected from the DFT calculations, the shift is quite small, and the fitting of the peak profile is not improved. Therefore, we believe that deviations in the peak shape arise from other effects, such as the neglect of anisotropy or tiny interchain couplings, inevitably present in a real material. The anisotropy could be larger than in usual cuprates because of the small energy difference between the Cu($3d_{x^2-y^2}$) and Cu($3d_{z^2-r^2}$) orbitals (see Fig.~\ref{lda}). At this point, we refrain from the numerical treatment of these effects and leave this difficult problem to future studies. 

\subsection{\label{sec:mH} High-field magnetization}
\begin{figure}[tb] \includegraphics[width=8.6cm]{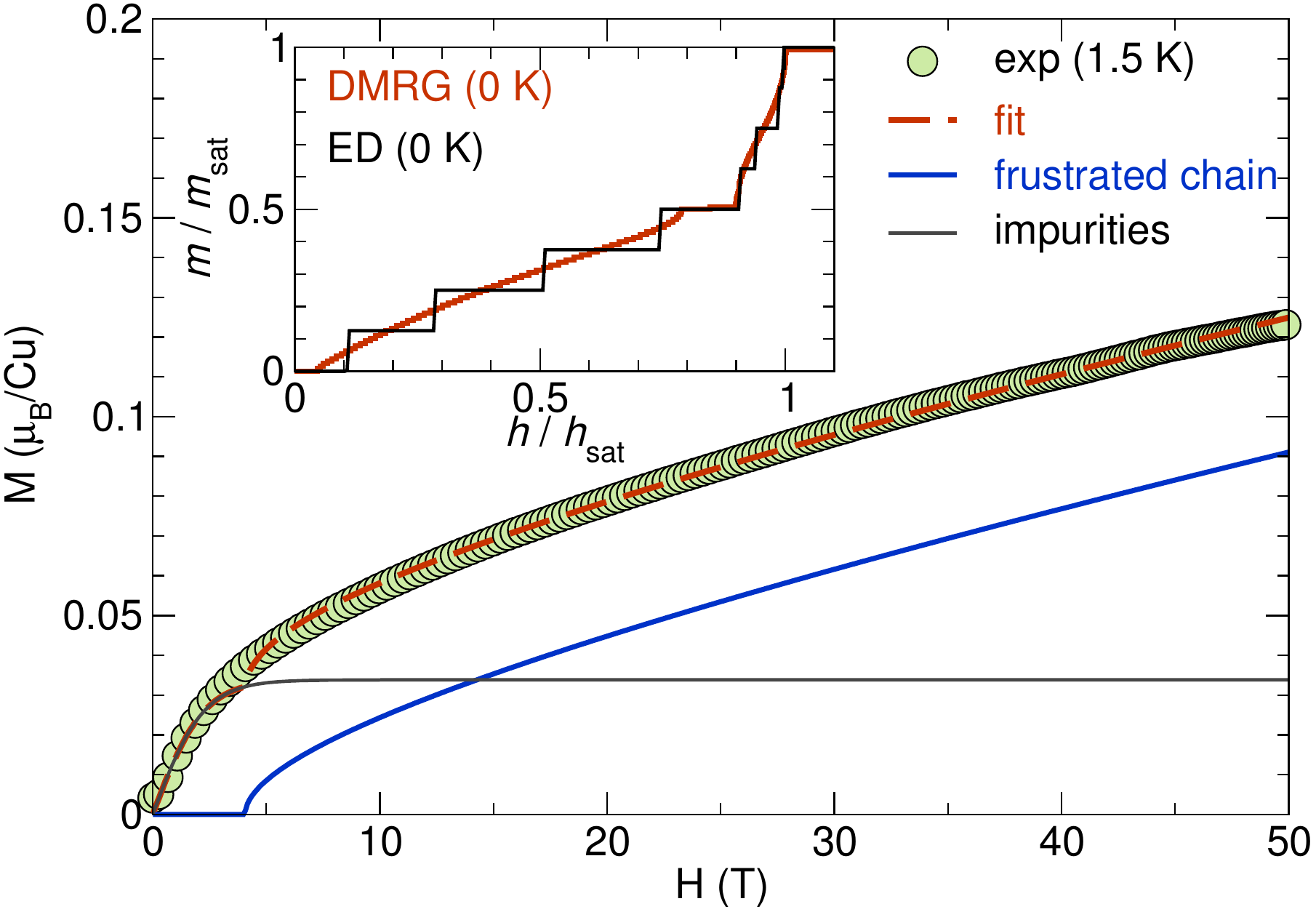}
\caption{\label{mH}(Color online) The experimental high-field magnetization curve (circles) and the fit (dashed line) with the sum of a Brillouin function (grey line), representing paramagnetic impurities, and a phenomenological function (Eq.~(\ref{Eq:mH}), blue line) for $h>\Delta$ that mimics the intrinsic magnetization of a frustrated $J_1-J_2-J_2'$ chain. Inset: the ground state magnetization of a frustrated $J_1-J_2-J_2'$ chain for $J_2:J_1$\,=\,0.5 and $J_2':J_2$\,=\,0.75 simulated using DMRG for $N$\,=\,256 spins (red line) and ED for $N$\,=\,16 spins (black line).}
\end{figure}

The high-field magnetization curve $M(H)$ (Fig.~\ref{mH}) generally agrees with the literature data~(Fig.~4 in Ref.~\onlinecite{szen_1}). Two different regimes can be distinguished: in low fields up to ~4\,T, $M(H)$ shows a steep increase followed by a more gentle increase up to the highest measured field (50\,T). The low-field part is dominated by an extrinsic contribution originating from orphan spins and other $S$\,=\,$\frac12$ impurities. Their magnetization is described by the Brillouin function:
\begin{equation}
M^{\text{imp}}(H) = n\tanh\left({\frac{g\mu_{\text{B}}}{2k_{\text{B}}}H}\right),
\end{equation}
where $n$ is the impurity content. The intrinsic contribution originates solely from the $J_1-J_2-J_2'$ chains, because the side chains remain in the singlet state over the whole range of accessible magnetic fields. 

The $M(H)$ curves calculated from ED and DMRG are shown in the inset of Fig.~\ref{mH}. After a narrow flat region associated with the spin gap, the magnetization starts increasing up to $h/h_{\rm sat}\simeq 0.8$ ($h\simeq 101$\,T assuming $J_1=68$\,K from the susceptibility fit), which is well above the field range of our experiment. Therefore, the only experimentally traceable feature of this curve is the closing of the spin gap in low fields. We thus used a phenomenological expression:
\begin{equation}
M^{\text{chain}}(H) = \frac{1+\text{sgn}(\tilde{H})}{2}\left(\alpha\tilde{H} + \beta\sqrt{\tilde{H}}\right),
\end{equation}
where $\tilde{H} = H-\Delta/g\mu_{\text{B}}$, $\Delta$ is the spin gap, and $\alpha$ and $\beta$ are free parameters which depend on $J_2'/J_2$ and $J_2/J_1$ ratios. Hence we can add $M^{\text{imp}}$ and describe the experimental curve as a sum of two contributions:
\begin{equation}
\label{Eq:mH}
M(H) = M^{\text{chain}}(H, \alpha, \beta, \Delta) + M^{\text{imp}}(H, n).
\end{equation}

This way, we obtain a good agreement with the experiment for $n$\,=\,3.4\,\% (compare to 3.6\% from the susceptibility fit), $\alpha$\,=\,7.9$\times$10$^{-4}$, $\beta$\,=\,8.1$\times$10$^{-3}$, and $\Delta$\,=\,4.1\,K (Fig.~\ref{mH}). The only notable discrepancy between our model and the experiment is the cusp at $H$\,=\,$\Delta/g\mu_{\text{B}}$, which is not seen in the experimental curve due to thermal broadening. 

\section{Discussion}
\label{sec:discussion}
The rare Cu-mineral szenicsite, Cu$_3$(MoO$_4$)(OH)$_4$, has been discussed in the present paper, and its microscopic magnetic model has been derived. At first glance, the triple-chains of Cu$^{2+}$ ions in the crystal structure might be considered as 3-leg spin ladders. Our calculations of the exchange couplings and subsequent modeling of the experimental data, however, reveal that side chains split into simple dimers with a strong AFM coupling ($>200$\,K), which govern the high-temperature behavior. At low temperatures, the system can be mapped onto the AFM \jj~model with the weak alternation of next-nearest-neighbor couplings ($J_2$--$J_2'$). 

\begin{figure}[tb] \includegraphics[width=8.6cm]{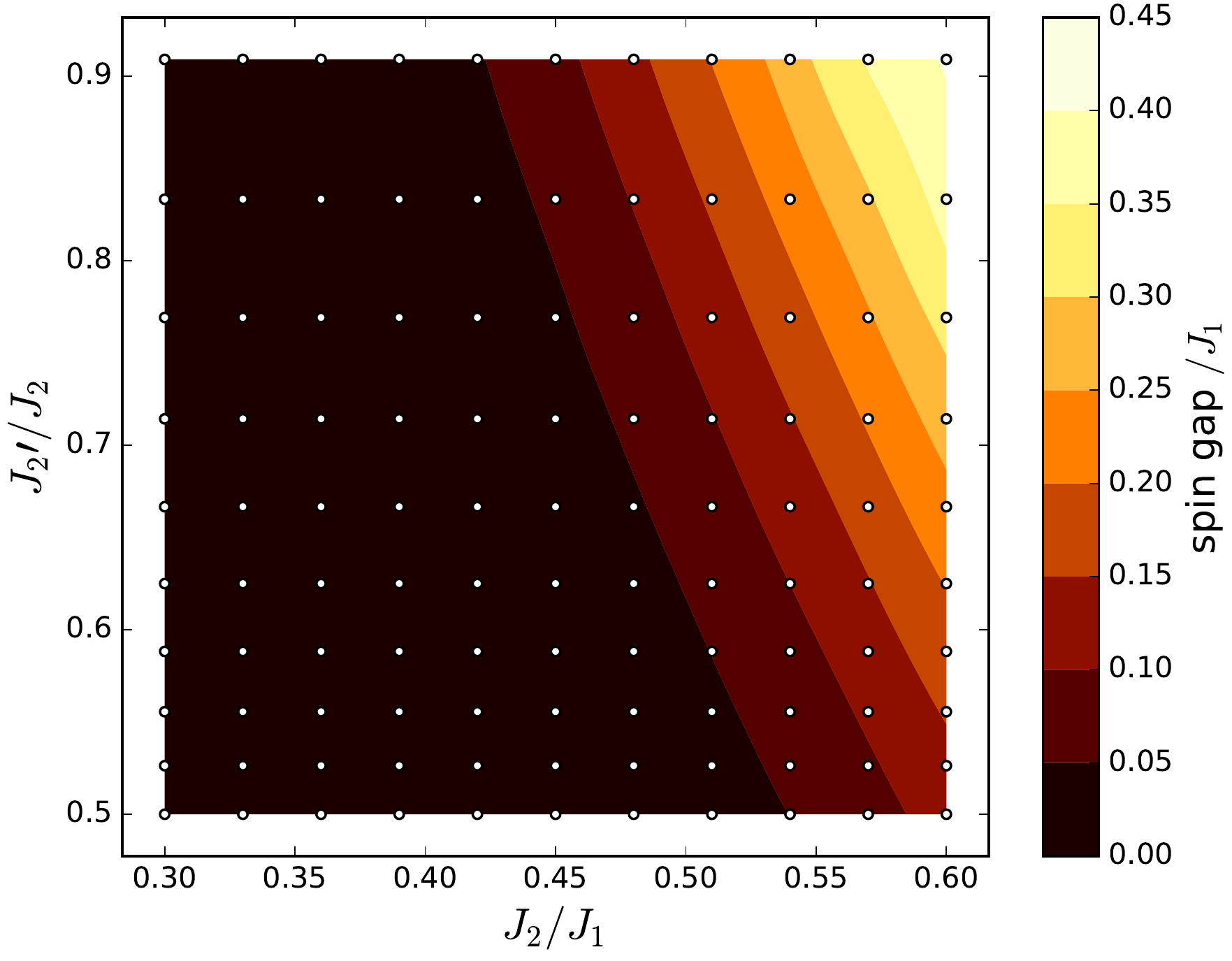}
\caption{\label{gap}(Color online) DMRG simulations of the spin gap as a function of the two ratios $J_2/J_1$ and $J_2'/J_2$ for the AFM $J_1$-$J_2$-$J_2'$ zigzag chain. The spin gap $\Delta \simeq 0.06/J_1$.}
\end{figure}

From DFT+$U$ calculations we obtained the ratio $J_2/J_1\approx 0.5$. Experimentally, this ratio is tracked by the position of the low-temperature peak in $C_{\rm mag}/T$ resulting in $J_2/J_1\simeq 0.45$, although the peak shape is not well reproduced. The spin gap is also sensitive to the $J_2/J_1$ ratio~\cite{white1996}. Our magnetization data are consistent with a small spin gap of 4.1\,K, about 6\% of $J_1$ assuming $J_1=68$\,K (see Table~\ref{sus_fit}). In Fig.~\ref{gap}, we show the spin gap of the $J_1-J_2-J_2'$ frustrated spin chain calculated with DMRG. The spin gap $\Delta\simeq 0.06J_1$ is compatible with $J_2/J_1=0.45$ and $J_2'/J_2\simeq 0.8$, in agreement with our TMRG analysis of the peak position in $C_{\text{mag}}/T$. 

\begin{figure}[tb] \includegraphics[width=8.6cm]{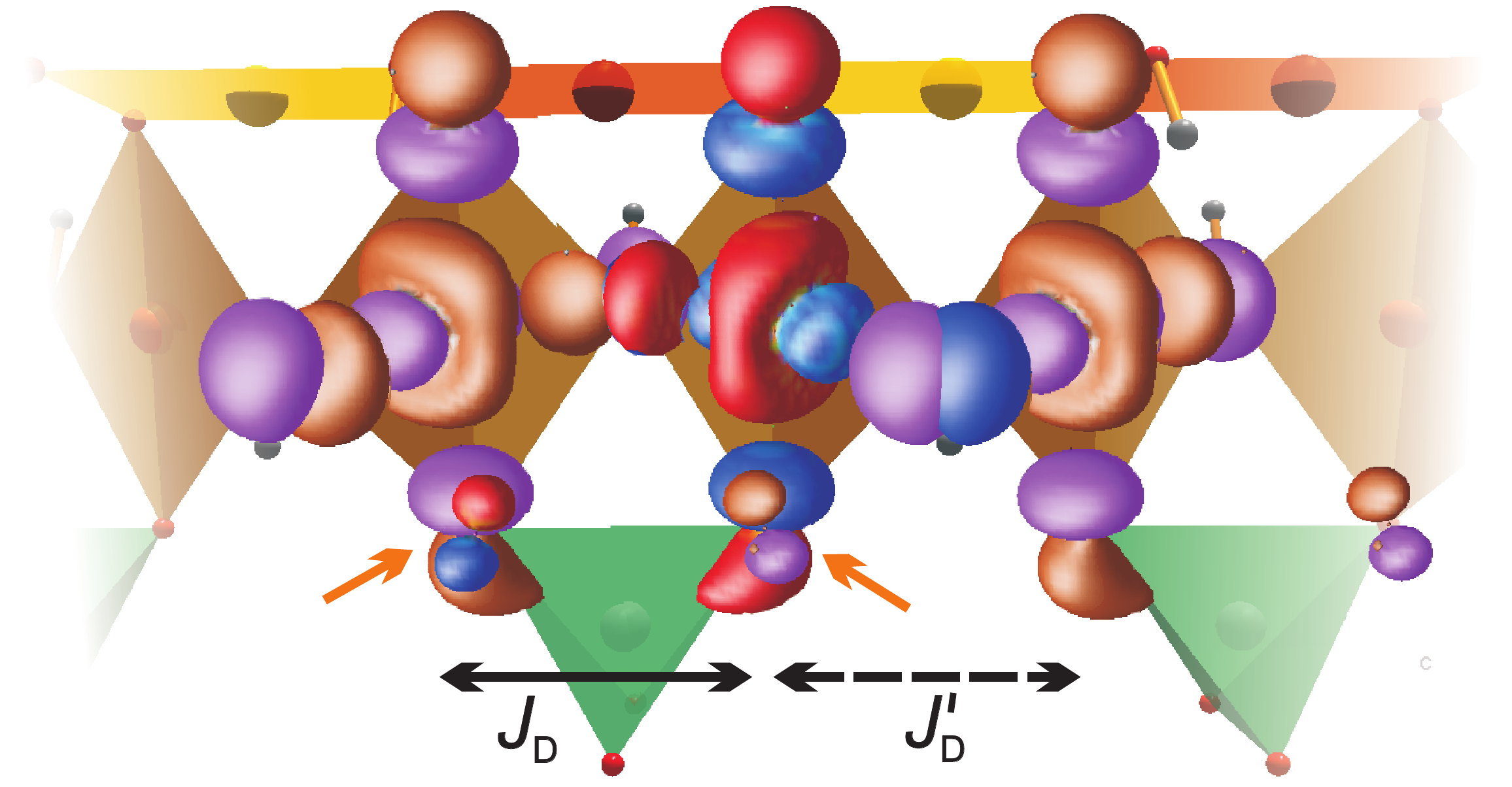}
\caption{\label{WF}(Color online) The Wannier functions on the Cu1 site. The MoO$_4$ tetrahedra are shown in green. The orange arrows indicate the sizable overlap of the Wannier functions at the MoO$_4$ tetrahedra. Only the $J_D$ exchange pathways involve the MoO$_4$ groups. Different colors for the Wannier functions (red/blue and lilac/orange) are used for illustrative reasons only. Different colors of a single Wannier function denote its different phases.}
\end{figure}

The manifestation of the AFM \jj~physics in szenicsite at low temperatures is essentially related to the formation of strongly coupled AFM dimers on the side chains. This could hardly be anticipated regarding the very similar bridging geometry for $J_D$ and $J_D'$. We will, thus, take a closer look at the details of these exchange pathways. In previous studies, we have pointed out the crucial importance of hydrogen atoms bonded to the bridging oxygen~\cite{malachite,clinoclase}. However, the O--H distances and Cu--O--H angles involved in $J_D$ and $J_D'$ differ only by 0.005~\r{A} and 1.95$^{\circ}$, respectively, which are way too small to account for the large difference in the coupling strengths. Another reason could be the MoO$_4$ groups bridging only the CuO$_4$ plaquettes along $J_D$ and not along $J_D'$ (Fig.~\ref{WF}). The WFs indeed exhibit sizable contributions from the MoO$_4$ groups, in particular O(2$p$) and small Mo(4$d$,5$s$) contributions, resulting in a considerably enhanced overlap along $J_D$, which is responsible for the large AFM coupling. 

The molybdate group, however, does not affect the coupling $J_D'$, and here hydrogen atoms come into play. The O--H bond is rotated by 50.5$^{\circ}$ out of the plane spanned by Cu1--O--Cu1, where O is the bridging oxygen in the side chain. This effect is responsible for the small $t_D'=-33$\,meV, in contrast with a much larger value expected for the Cu--O--Cu bridging angle of 103.4$^{\circ}$ based on the GKA rules~\cite{gka1,*gka2,*gka3}. By performing calculations for a hypothetical in-plane position of H, we obtained a much larger $t_D'=-135$\,meV indeed. The same argument applies to $t_D$, comprising a very similar out-of-plane position of H. This hopping would also increase to $t_D=-227$\,meV for the hypothetical in-plane position of H. Thus, both NN transfer integrals in the side chain increase by almost the same amount of about 100\,K when hydrogens are moved in-plane. This tendency of an out-of-plane position of H suppressing the transfer probability agrees with our previous studies on Cu$^{2+}$-minerals~\cite{callaghanite,clinoclase,langite}.

All other exchange couplings listed in Table~\ref{tJ} follow the GKA rules and are compatible with the size of the respective bridging angles (see Table~\ref{tJ}). The strength of the NN coupling $J_1$ is typical for the edge-sharing geometry with a bridging angle of about 100\degr~\cite{malachite,clinoclase}. Remarkably, though, such a high bridging angle has not been seen in the majority of the \jj-chain compounds, where the lower bridging angles of $90-95^{\circ}$ produce FM $J_1$. 

A recent nuclear magnetic resonance (NMR)-study on szencisite~\cite{nmr_szen} proposed a different spin lattice consisting of uniform antiferromagnetic chains without a spin gap. However, the respective susceptibility fit could not reproduce the experimental data below about 80\,K. The NMR spectra were measured at a magnetic field of 4\,T where, according to our calculations, the spin gap is already closed. Therefore, experimental observations of Ref.~\onlinecite{nmr_szen} do not contradict our model. 

While our thermodynamic measurements are consistent with the small spin gap expected in the \jj-model of szenicsite, experimental signatures of this spin gap are obscured by the impurity contribution that is very pronounced at low temperatures. Direct experimental probe of the spin gap is still open. NMR data at low fields may be useful in this respect, for they should be less affected by magnetic impurities than bulk properties, such as magnetization. Preparation of a synthetic sample would be interesting too and may help to observe the small spin gap even in thermodynamic measurements. Inelastic neutron scattering has sufficient energy resolution to probe the spin gap on the order of 4\,K, but a synthetic deuterated sample would be required.

The crystal structure of szenicsite is closely related to that of antlerite, Cu$_3$(SO$_4$)(OH)$_4$~\cite{xxstr}, which attracted interest due to the proposed "idle spin" behavior~\cite{antlerite_2003,antlerite_2007} and its rich field-induced magnetic phase diagram~\cite{antlerite_2011,*antlerite_2013}. Both compounds feature very similar \mbox{Cu--Cu} distances and Cu--O--Cu angles, so that one would expect similar magnetic properties as well. However, the magnetic properties of the two compounds are actually very different. First, antlerite undergoes an LRO transition below 5.3\,K~\cite{antlerite_2009}, which is not observed in szenicsite. Second, the Curie-Weiss fit provides a negative $\theta$ value of $-1.96$\,K~\cite{antlerite_2003}, implying sizable ferromagnetic interactions. For szenicsite $\theta$ could be evaluated to about 68\,K, and all couplings are AFM. 

The differences between szenicsite and antlerite are mostly related to the side chains. For the side chains in antlerite, both $J_D$ and $J_D'$ are reported as ferromagnetic~\cite{antlerite_2003,antlerite_2012}, so that spin dimers are not formed, and spins in the side chains develop long-range magnetic order. For the central chain, both szenicsite and antlerite seem to represent the AFM $J_1-J_2-J_2'$-model, and the idle-spin behavior (zero ordered moment in the central chain) is well accounted for by the spin gap formation in the $J_1-J_2-J_2'$-chain with AFM $J_1$. 

\section{Summary}
\label{sec:summary}
We reported magnetic behavior and microscopic magnetic model of the rare Cu$^{2+}$-mineral szenicsite whose crystal structure features isolated triple chains. These chains consist of a central chain formed by edge-sharing CuO$_4$ plaquettes and two side chains built of corner-sharing plaquettes with MoO$_4$-groups attached. A consistent magnetic model could be derived on the basis of LDA and LSDA+$U$ calculations of exchange coupling constants, fits of susceptibility data, and model analysis of the magnetization and heat capacity data. 

Even though the crystal structure suggests a 3-leg-ladder magnetic model, it turns out that the side chains can be reduced to AFM dimers with a coupling strength of about 200\,K, where the central chain features an AFM NN coupling of about 68\,K and alternating NNN couplings. The couplings between the chains are quite small, below 30\,K, and of minor importance for the magnetic model because of the frustration and different energy scales in the side chains versus the central chain. Accordingly, the low-temperature physics of szenicsite is well described by an original version of AFM \jj~model with alternating $J_2$--$J_2'$ couplings. 

According to the $J_2/J_1$ ratio of about 0.45, a small spin gap should be present in szenicsite. Its upper limit of about 4\,K is consistent with our magnetization measurements. A similar mechanism of gap opening is likely responsible for the formation of idle spins in the closely related mineral antlerite. However, the physics of side chains is largely different in the two minerals. We ascribed these differences to effects of side groups and hydrogen positions that play crucial role for magnetic interactions in cupric minerals.

\acknowledgments
We acknowledge experimental support by Yurii Prots and Horst Borrmann (laboratory XRD), Christoph Klausnitzer (low-temperature specific heat measurements), and Carolina Curfs (data collection at ID31). We are grateful to Yurii Skourskii for high-field magnetization measurements and acknowledge the support of the HLD at HZDR, member of EMFL. We would like to thank Prof. Klaus Thalheim and the Senckenberg Naturhistorische Sammlung Dresden for providing the szenicsite sample from Chile. OJ was partly supported by the European Research Council under the European Unions Seventh Framework Program FP7/ERC through grant agreement n.306447. AT was supported by Federal Ministry for Education and Research through the Sofja Kovalevskaya Award of Alexander von Humboldt Foundation. The German Research Foundation (SFB 1143) of the Deutsche Forschungsgemeinschaft supported SN.


\begin{thebibliography}{67}%
\makeatletter
\providecommand \@ifxundefined [1]{%
 \@ifx{#1\undefined}
}%
\providecommand \@ifnum [1]{%
 \ifnum #1\expandafter \@firstoftwo
 \else \expandafter \@secondoftwo
 \fi
}%
\providecommand \@ifx [1]{%
 \ifx #1\expandafter \@firstoftwo
 \else \expandafter \@secondoftwo
 \fi
}%
\providecommand \natexlab [1]{#1}%
\providecommand \enquote  [1]{``#1''}%
\providecommand \bibnamefont  [1]{#1}%
\providecommand \bibfnamefont [1]{#1}%
\providecommand \citenamefont [1]{#1}%
\providecommand \href@noop [0]{\@secondoftwo}%
\providecommand \href [0]{\begingroup \@sanitize@url \@href}%
\providecommand \@href[1]{\@@startlink{#1}\@@href}%
\providecommand \@@href[1]{\endgroup#1\@@endlink}%
\providecommand \@sanitize@url [0]{\catcode `\\12\catcode `\$12\catcode
  `\&12\catcode `\#12\catcode `\^12\catcode `\_12\catcode `\%12\relax}%
\providecommand \@@startlink[1]{}%
\providecommand \@@endlink[0]{}%
\providecommand \url  [0]{\begingroup\@sanitize@url \@url }%
\providecommand \@url [1]{\endgroup\@href {#1}{\urlprefix }}%
\providecommand \urlprefix  [0]{URL }%
\providecommand \Eprint [0]{\href }%
\providecommand \doibase [0]{http://dx.doi.org/}%
\providecommand \selectlanguage [0]{\@gobble}%
\providecommand \bibinfo  [0]{\@secondoftwo}%
\providecommand \bibfield  [0]{\@secondoftwo}%
\providecommand \translation [1]{[#1]}%
\providecommand \BibitemOpen [0]{}%
\providecommand \bibitemStop [0]{}%
\providecommand \bibitemNoStop [0]{.\EOS\space}%
\providecommand \EOS [0]{\spacefactor3000\relax}%
\providecommand \BibitemShut  [1]{\csname bibitem#1\endcsname}%
\let\auto@bib@innerbib\@empty
\bibitem [{\citenamefont {Mikeska}\ and\ \citenamefont
  {Kolezhuk}(2004)}]{1d_mag}%
  \BibitemOpen
  \bibfield  {author} {\bibinfo {author} {\bibfnamefont {H.-J.}\ \bibnamefont
  {Mikeska}}\ and\ \bibinfo {author} {\bibfnamefont {A.~K.}\ \bibnamefont
  {Kolezhuk}},\ }\href {\doibase 10.1007/BFb0119591} {\bibfield  {journal}
  {\bibinfo  {journal} {Lect. Notes Phys.}\ }\textbf {\bibinfo {volume}
  {645}},\ \bibinfo {pages} {1} (\bibinfo {year} {2004})}\BibitemShut {NoStop}%
\bibitem [{\citenamefont {Parkinson}\ and\ \citenamefont
  {Farnell}(2010)}]{quant_magn}%
  \BibitemOpen
  \bibfield  {author} {\bibinfo {author} {\bibfnamefont {J.~B.}\ \bibnamefont
  {Parkinson}}\ and\ \bibinfo {author} {\bibfnamefont {D.~J.}\ \bibnamefont
  {Farnell}},\ }\href {\doibase 10.1007/978-3-642-13290-2_11} {\bibfield
  {journal} {\bibinfo  {journal} {Lect. Notes Phys.}\ }\textbf {\bibinfo
  {volume} {816}},\ \bibinfo {pages} {135} (\bibinfo {year}
  {2010})}\BibitemShut {NoStop}%
\bibitem [{\citenamefont {Chitra}\ \emph {et~al.}(1995)\citenamefont {Chitra},
  \citenamefont {Pati}, \citenamefont {Krishnamurthy}, \citenamefont {Sen},\
  and\ \citenamefont {Ramasesha}}]{chitra1995}%
  \BibitemOpen
  \bibfield  {author} {\bibinfo {author} {\bibfnamefont {R.}~\bibnamefont
  {Chitra}}, \bibinfo {author} {\bibfnamefont {S.}~\bibnamefont {Pati}},
  \bibinfo {author} {\bibfnamefont {H.~R.}\ \bibnamefont {Krishnamurthy}},
  \bibinfo {author} {\bibfnamefont {D.}~\bibnamefont {Sen}}, \ and\ \bibinfo
  {author} {\bibfnamefont {S.}~\bibnamefont {Ramasesha}},\ }\href@noop {}
  {\bibfield  {journal} {\bibinfo  {journal} {Phys. Rev. B}\ }\textbf {\bibinfo
  {volume} {52}},\ \bibinfo {pages} {6581} (\bibinfo {year} {1995})},\ \bibinfo
  {note} {cond-mat/9412016}\BibitemShut {NoStop}%
\bibitem [{\citenamefont {Sudan}\ \emph {et~al.}(2009)\citenamefont {Sudan},
  \citenamefont {L\"uscher},\ and\ \citenamefont {L\"auchli}}]{sudan2009}%
  \BibitemOpen
  \bibfield  {author} {\bibinfo {author} {\bibfnamefont {J.}~\bibnamefont
  {Sudan}}, \bibinfo {author} {\bibfnamefont {A.}~\bibnamefont {L\"uscher}}, \
  and\ \bibinfo {author} {\bibfnamefont {A.~M.}\ \bibnamefont {L\"auchli}},\
  }\href@noop {} {\bibfield  {journal} {\bibinfo  {journal} {Phys. Rev. B}\
  }\textbf {\bibinfo {volume} {80}},\ \bibinfo {pages} {140402(R)} (\bibinfo
  {year} {2009})}\BibitemShut {NoStop}%
\bibitem [{\citenamefont {Zinke}\ \emph {et~al.}(2009)\citenamefont {Zinke},
  \citenamefont {Drechsler},\ and\ \citenamefont {Richter}}]{zinke2009}%
  \BibitemOpen
  \bibfield  {author} {\bibinfo {author} {\bibfnamefont {R.}~\bibnamefont
  {Zinke}}, \bibinfo {author} {\bibfnamefont {S.-L.}\ \bibnamefont
  {Drechsler}}, \ and\ \bibinfo {author} {\bibfnamefont {J.}~\bibnamefont
  {Richter}},\ }\href@noop {} {\bibfield  {journal} {\bibinfo  {journal} {Phys.
  Rev. B}\ }\textbf {\bibinfo {volume} {79}},\ \bibinfo {pages} {094425}
  (\bibinfo {year} {2009})},\ \bibinfo {note} {arXiv:0807.3431}\BibitemShut
  {NoStop}%
\bibitem [{\citenamefont {Furukawa}\ \emph {et~al.}(2010)\citenamefont
  {Furukawa}, \citenamefont {Sato},\ and\ \citenamefont
  {Onoda}}]{furukawa2010}%
  \BibitemOpen
  \bibfield  {author} {\bibinfo {author} {\bibfnamefont {S.}~\bibnamefont
  {Furukawa}}, \bibinfo {author} {\bibfnamefont {M.}~\bibnamefont {Sato}}, \
  and\ \bibinfo {author} {\bibfnamefont {S.}~\bibnamefont {Onoda}},\
  }\href@noop {} {\bibfield  {journal} {\bibinfo  {journal} {Phys. Rev. Lett.}\
  }\textbf {\bibinfo {volume} {105}},\ \bibinfo {pages} {257205} (\bibinfo
  {year} {2010})},\ \bibinfo {note} {arXiv:1003.3940}\BibitemShut {NoStop}%
\bibitem [{\citenamefont {Sirker}\ \emph {et~al.}(2011)\citenamefont {Sirker},
  \citenamefont {Krivnov}, \citenamefont {Dmitriev}, \citenamefont {Herzog},
  \citenamefont {Janson}, \citenamefont {Nishimoto}, \citenamefont
  {Drechsler},\ and\ \citenamefont {Richter}}]{sirker2011}%
  \BibitemOpen
  \bibfield  {author} {\bibinfo {author} {\bibfnamefont {J.}~\bibnamefont
  {Sirker}}, \bibinfo {author} {\bibfnamefont {V.~Y.}\ \bibnamefont {Krivnov}},
  \bibinfo {author} {\bibfnamefont {D.~V.}\ \bibnamefont {Dmitriev}}, \bibinfo
  {author} {\bibfnamefont {A.}~\bibnamefont {Herzog}}, \bibinfo {author}
  {\bibfnamefont {O.}~\bibnamefont {Janson}}, \bibinfo {author} {\bibfnamefont
  {S.}~\bibnamefont {Nishimoto}}, \bibinfo {author} {\bibfnamefont {S.-L.}\
  \bibnamefont {Drechsler}}, \ and\ \bibinfo {author} {\bibfnamefont
  {J.}~\bibnamefont {Richter}},\ }\href@noop {} {\bibfield  {journal} {\bibinfo
   {journal} {Phys. Rev. B}\ }\textbf {\bibinfo {volume} {84}},\ \bibinfo
  {pages} {144403} (\bibinfo {year} {2011})}\BibitemShut {NoStop}%
\bibitem [{\citenamefont {Balents}\ and\ \citenamefont
  {Starykh}(2016)}]{balents2016}%
  \BibitemOpen
  \bibfield  {author} {\bibinfo {author} {\bibfnamefont {L.}~\bibnamefont
  {Balents}}\ and\ \bibinfo {author} {\bibfnamefont {O.~A.}\ \bibnamefont
  {Starykh}},\ }\href@noop {} {\bibfield  {journal} {\bibinfo  {journal} {Phys.
  Rev. Lett.}\ }\textbf {\bibinfo {volume} {116}},\ \bibinfo {pages} {177201}
  (\bibinfo {year} {2016})}\BibitemShut {NoStop}%
\bibitem [{\citenamefont {Banks}\ \emph {et~al.}(2009)\citenamefont {Banks},
  \citenamefont {Kremer}, \citenamefont {Hoch}, \citenamefont {Simon},
  \citenamefont {Ouladdiaf}, \citenamefont {Broto}, \citenamefont {Rakoto},
  \citenamefont {Lee},\ and\ \citenamefont {Whangbo}}]{banks2009}%
  \BibitemOpen
  \bibfield  {author} {\bibinfo {author} {\bibfnamefont {M.~G.}\ \bibnamefont
  {Banks}}, \bibinfo {author} {\bibfnamefont {R.~K.}\ \bibnamefont {Kremer}},
  \bibinfo {author} {\bibfnamefont {C.}~\bibnamefont {Hoch}}, \bibinfo {author}
  {\bibfnamefont {A.}~\bibnamefont {Simon}}, \bibinfo {author} {\bibfnamefont
  {B.}~\bibnamefont {Ouladdiaf}}, \bibinfo {author} {\bibfnamefont {J.-M.}\
  \bibnamefont {Broto}}, \bibinfo {author} {\bibfnamefont {H.}~\bibnamefont
  {Rakoto}}, \bibinfo {author} {\bibfnamefont {C.}~\bibnamefont {Lee}}, \ and\
  \bibinfo {author} {\bibfnamefont {M.-H.}\ \bibnamefont {Whangbo}},\
  }\href@noop {} {\bibfield  {journal} {\bibinfo  {journal} {Phys. Rev. B}\
  }\textbf {\bibinfo {volume} {80}},\ \bibinfo {pages} {024404} (\bibinfo
  {year} {2009})},\ \bibinfo {note} {arXiv:0904.2929}\BibitemShut {NoStop}%
\bibitem [{\citenamefont {Lee}\ \emph {et~al.}(2012)\citenamefont {Lee},
  \citenamefont {Liu}, \citenamefont {Whangbo}, \citenamefont {Koo},
  \citenamefont {Kremer},\ and\ \citenamefont {Simon}}]{lee2012}%
  \BibitemOpen
  \bibfield  {author} {\bibinfo {author} {\bibfnamefont {C.}~\bibnamefont
  {Lee}}, \bibinfo {author} {\bibfnamefont {J.}~\bibnamefont {Liu}}, \bibinfo
  {author} {\bibfnamefont {M.-H.}\ \bibnamefont {Whangbo}}, \bibinfo {author}
  {\bibfnamefont {H.-J.}\ \bibnamefont {Koo}}, \bibinfo {author} {\bibfnamefont
  {R.~K.}\ \bibnamefont {Kremer}}, \ and\ \bibinfo {author} {\bibfnamefont
  {A.}~\bibnamefont {Simon}},\ }\href@noop {} {\bibfield  {journal} {\bibinfo
  {journal} {Phys. Rev. B}\ }\textbf {\bibinfo {volume} {86}},\ \bibinfo
  {pages} {060407(R)} (\bibinfo {year} {2012})}\BibitemShut {NoStop}%
\bibitem [{\citenamefont {Willenberg}\ \emph {et~al.}(2012)\citenamefont
  {Willenberg}, \citenamefont {Sch\"apers}, \citenamefont {Rule}, \citenamefont
  {S\"ullow}, \citenamefont {Reehuis}, \citenamefont {Ryll}, \citenamefont
  {Klemke}, \citenamefont {Kiefer}, \citenamefont {Schottenhamel},
  \citenamefont {B\"uchner}, \citenamefont {Ouladdiaf}, \citenamefont {Uhlarz},
  \citenamefont {Beyer}, \citenamefont {Wosnitza},\ and\ \citenamefont
  {Wolter}}]{willenberg2012}%
  \BibitemOpen
  \bibfield  {author} {\bibinfo {author} {\bibfnamefont {B.}~\bibnamefont
  {Willenberg}}, \bibinfo {author} {\bibfnamefont {M.}~\bibnamefont
  {Sch\"apers}}, \bibinfo {author} {\bibfnamefont {K.~C.}\ \bibnamefont
  {Rule}}, \bibinfo {author} {\bibfnamefont {S.}~\bibnamefont {S\"ullow}},
  \bibinfo {author} {\bibfnamefont {M.}~\bibnamefont {Reehuis}}, \bibinfo
  {author} {\bibfnamefont {H.}~\bibnamefont {Ryll}}, \bibinfo {author}
  {\bibfnamefont {B.}~\bibnamefont {Klemke}}, \bibinfo {author} {\bibfnamefont
  {K.}~\bibnamefont {Kiefer}}, \bibinfo {author} {\bibfnamefont
  {W.}~\bibnamefont {Schottenhamel}}, \bibinfo {author} {\bibfnamefont
  {B.}~\bibnamefont {B\"uchner}}, \bibinfo {author} {\bibfnamefont
  {B.}~\bibnamefont {Ouladdiaf}}, \bibinfo {author} {\bibfnamefont
  {M.}~\bibnamefont {Uhlarz}}, \bibinfo {author} {\bibfnamefont
  {R.}~\bibnamefont {Beyer}}, \bibinfo {author} {\bibfnamefont
  {J.}~\bibnamefont {Wosnitza}}, \ and\ \bibinfo {author} {\bibfnamefont
  {A.~U.~B.}\ \bibnamefont {Wolter}},\ }\href@noop {} {\bibfield  {journal}
  {\bibinfo  {journal} {Phys. Rev. Lett.}\ }\textbf {\bibinfo {volume} {108}},\
  \bibinfo {pages} {117202} (\bibinfo {year} {2012})}\BibitemShut {NoStop}%
\bibitem [{\citenamefont {Pregelj}\ \emph {et~al.}(2015)\citenamefont
  {Pregelj}, \citenamefont {Zorko}, \citenamefont {Zaharko}, \citenamefont
  {Nojiri}, \citenamefont {Berger}, \citenamefont {Chapon},\ and\ \citenamefont
  {Ar{\u c}on}}]{pregelj2015}%
  \BibitemOpen
  \bibfield  {author} {\bibinfo {author} {\bibfnamefont {M.}~\bibnamefont
  {Pregelj}}, \bibinfo {author} {\bibfnamefont {A.}~\bibnamefont {Zorko}},
  \bibinfo {author} {\bibfnamefont {O.}~\bibnamefont {Zaharko}}, \bibinfo
  {author} {\bibfnamefont {H.}~\bibnamefont {Nojiri}}, \bibinfo {author}
  {\bibfnamefont {H.}~\bibnamefont {Berger}}, \bibinfo {author} {\bibfnamefont
  {L.~C.}\ \bibnamefont {Chapon}}, \ and\ \bibinfo {author} {\bibfnamefont
  {D.}~\bibnamefont {Ar{\u c}on}},\ }\href@noop {} {\bibfield  {journal}
  {\bibinfo  {journal} {Nature Comm.}\ }\textbf {\bibinfo {volume} {6}},\
  \bibinfo {pages} {7255} (\bibinfo {year} {2015})}\BibitemShut {NoStop}%
\bibitem [{\citenamefont {Park}\ \emph {et~al.}(2007)\citenamefont {Park},
  \citenamefont {Choi}, \citenamefont {Zhang},\ and\ \citenamefont
  {Cheong}}]{park2007}%
  \BibitemOpen
  \bibfield  {author} {\bibinfo {author} {\bibfnamefont {S.}~\bibnamefont
  {Park}}, \bibinfo {author} {\bibfnamefont {Y.~J.}\ \bibnamefont {Choi}},
  \bibinfo {author} {\bibfnamefont {C.~L.}\ \bibnamefont {Zhang}}, \ and\
  \bibinfo {author} {\bibfnamefont {S.-W.}\ \bibnamefont {Cheong}},\
  }\href@noop {} {\bibfield  {journal} {\bibinfo  {journal} {Phys. Rev. Lett.}\
  }\textbf {\bibinfo {volume} {98}},\ \bibinfo {pages} {057601} (\bibinfo
  {year} {2007})}\BibitemShut {NoStop}%
\bibitem [{\citenamefont {Naito}\ \emph {et~al.}(2007)\citenamefont {Naito},
  \citenamefont {Sato}, \citenamefont {Yasui}, \citenamefont {Kobayashi},
  \citenamefont {Kobayashi},\ and\ \citenamefont {Sato}}]{naito2007}%
  \BibitemOpen
  \bibfield  {author} {\bibinfo {author} {\bibfnamefont {Y.}~\bibnamefont
  {Naito}}, \bibinfo {author} {\bibfnamefont {K.}~\bibnamefont {Sato}},
  \bibinfo {author} {\bibfnamefont {Y.}~\bibnamefont {Yasui}}, \bibinfo
  {author} {\bibfnamefont {Y.}~\bibnamefont {Kobayashi}}, \bibinfo {author}
  {\bibfnamefont {Y.}~\bibnamefont {Kobayashi}}, \ and\ \bibinfo {author}
  {\bibfnamefont {M.}~\bibnamefont {Sato}},\ }\href@noop {} {\bibfield
  {journal} {\bibinfo  {journal} {J. Phys. Soc. Jpn.}\ }\textbf {\bibinfo
  {volume} {76}},\ \bibinfo {pages} {023708} (\bibinfo {year} {2007})},\
  \bibinfo {note} {cond-mat/0611659}\BibitemShut {NoStop}%
\bibitem [{\citenamefont {Schrettle}\ \emph {et~al.}(2008)\citenamefont
  {Schrettle}, \citenamefont {Krohns}, \citenamefont {Lunkenheimer},
  \citenamefont {Hemberger}, \citenamefont {B\"uttgen}, \citenamefont {{Krug
  von Nidda}}, \citenamefont {Prokofiev},\ and\ \citenamefont
  {Loidl}}]{schrettle2008}%
  \BibitemOpen
  \bibfield  {author} {\bibinfo {author} {\bibfnamefont {F.}~\bibnamefont
  {Schrettle}}, \bibinfo {author} {\bibfnamefont {S.}~\bibnamefont {Krohns}},
  \bibinfo {author} {\bibfnamefont {P.}~\bibnamefont {Lunkenheimer}}, \bibinfo
  {author} {\bibfnamefont {J.}~\bibnamefont {Hemberger}}, \bibinfo {author}
  {\bibfnamefont {N.}~\bibnamefont {B\"uttgen}}, \bibinfo {author}
  {\bibfnamefont {H.-A.}\ \bibnamefont {{Krug von Nidda}}}, \bibinfo {author}
  {\bibfnamefont {A.~V.}\ \bibnamefont {Prokofiev}}, \ and\ \bibinfo {author}
  {\bibfnamefont {A.}~\bibnamefont {Loidl}},\ }\href@noop {} {\bibfield
  {journal} {\bibinfo  {journal} {Phys. Rev. B}\ }\textbf {\bibinfo {volume}
  {77}},\ \bibinfo {pages} {144101} (\bibinfo {year} {2008})}\BibitemShut
  {NoStop}%
\bibitem [{\citenamefont {Seki}\ \emph {et~al.}(2010)\citenamefont {Seki},
  \citenamefont {Kurumaji}, \citenamefont {Ishiwata}, \citenamefont {Matsui},
  \citenamefont {Murakawa}, \citenamefont {Tokunaga}, \citenamefont {Kaneko},
  \citenamefont {Hasegawa},\ and\ \citenamefont {Tokura}}]{seki2010}%
  \BibitemOpen
  \bibfield  {author} {\bibinfo {author} {\bibfnamefont {S.}~\bibnamefont
  {Seki}}, \bibinfo {author} {\bibfnamefont {T.}~\bibnamefont {Kurumaji}},
  \bibinfo {author} {\bibfnamefont {S.}~\bibnamefont {Ishiwata}}, \bibinfo
  {author} {\bibfnamefont {H.}~\bibnamefont {Matsui}}, \bibinfo {author}
  {\bibfnamefont {H.}~\bibnamefont {Murakawa}}, \bibinfo {author}
  {\bibfnamefont {Y.}~\bibnamefont {Tokunaga}}, \bibinfo {author}
  {\bibfnamefont {Y.}~\bibnamefont {Kaneko}}, \bibinfo {author} {\bibfnamefont
  {T.}~\bibnamefont {Hasegawa}}, \ and\ \bibinfo {author} {\bibfnamefont
  {Y.}~\bibnamefont {Tokura}},\ }\href@noop {} {\bibfield  {journal} {\bibinfo
  {journal} {Phys. Rev. B}\ }\textbf {\bibinfo {volume} {82}},\ \bibinfo
  {pages} {064424} (\bibinfo {year} {2010})},\ \bibinfo {note}
  {arXiv:1008.5226}\BibitemShut {NoStop}%
\bibitem [{\citenamefont {Yasui}\ \emph {et~al.}(2011)\citenamefont {Yasui},
  \citenamefont {Sato},\ and\ \citenamefont {Terasaki}}]{yasui2011}%
  \BibitemOpen
  \bibfield  {author} {\bibinfo {author} {\bibfnamefont {Y.}~\bibnamefont
  {Yasui}}, \bibinfo {author} {\bibfnamefont {M.}~\bibnamefont {Sato}}, \ and\
  \bibinfo {author} {\bibfnamefont {I.}~\bibnamefont {Terasaki}},\ }\href@noop
  {} {\bibfield  {journal} {\bibinfo  {journal} {J. Phys. Soc. Jpn.}\ }\textbf
  {\bibinfo {volume} {80}},\ \bibinfo {pages} {033707} (\bibinfo {year}
  {2011})}\BibitemShut {NoStop}%
\bibitem [{\citenamefont {Zhao}\ \emph {et~al.}(2012)\citenamefont {Zhao},
  \citenamefont {Hung}, \citenamefont {Li}, \citenamefont {Chen}, \citenamefont
  {Wu}, \citenamefont {Kremer}, \citenamefont {Banks}, \citenamefont {Simon},
  \citenamefont {Whangbo}, \citenamefont {Lee}, \citenamefont {Kim},
  \citenamefont {Kim},\ and\ \citenamefont {Kim}}]{zhao2012}%
  \BibitemOpen
  \bibfield  {author} {\bibinfo {author} {\bibfnamefont {L.}~\bibnamefont
  {Zhao}}, \bibinfo {author} {\bibfnamefont {T.-L.}\ \bibnamefont {Hung}},
  \bibinfo {author} {\bibfnamefont {C.-C.}\ \bibnamefont {Li}}, \bibinfo
  {author} {\bibfnamefont {Y.-Y.}\ \bibnamefont {Chen}}, \bibinfo {author}
  {\bibfnamefont {M.-K.}\ \bibnamefont {Wu}}, \bibinfo {author} {\bibfnamefont
  {R.~K.}\ \bibnamefont {Kremer}}, \bibinfo {author} {\bibfnamefont {M.~G.}\
  \bibnamefont {Banks}}, \bibinfo {author} {\bibfnamefont {A.}~\bibnamefont
  {Simon}}, \bibinfo {author} {\bibfnamefont {M.-H.}\ \bibnamefont {Whangbo}},
  \bibinfo {author} {\bibfnamefont {C.}~\bibnamefont {Lee}}, \bibinfo {author}
  {\bibfnamefont {J.~S.}\ \bibnamefont {Kim}}, \bibinfo {author} {\bibfnamefont
  {I.}~\bibnamefont {Kim}}, \ and\ \bibinfo {author} {\bibfnamefont {K.~H.}\
  \bibnamefont {Kim}},\ }\href@noop {} {\bibfield  {journal} {\bibinfo
  {journal} {Adv. Mater.}\ }\textbf {\bibinfo {volume} {24}},\ \bibinfo {pages}
  {2469} (\bibinfo {year} {2012})}\BibitemShut {NoStop}%
\bibitem [{\citenamefont {Kecke}\ \emph {et~al.}(2007)\citenamefont {Kecke},
  \citenamefont {Momoi},\ and\ \citenamefont {Furusaki}}]{j1j2_H1}%
  \BibitemOpen
  \bibfield  {author} {\bibinfo {author} {\bibfnamefont {L.}~\bibnamefont
  {Kecke}}, \bibinfo {author} {\bibfnamefont {T.}~\bibnamefont {Momoi}}, \ and\
  \bibinfo {author} {\bibfnamefont {A.}~\bibnamefont {Furusaki}},\ }\href
  {\doibase 10.1103/PhysRevB.76.060407} {\bibfield  {journal} {\bibinfo
  {journal} {Phys. Rev. B}\ }\textbf {\bibinfo {volume} {76}},\ \bibinfo
  {pages} {060407} (\bibinfo {year} {2007})}\BibitemShut {NoStop}%
\bibitem [{\citenamefont {Hikihara}\ \emph {et~al.}(2008)\citenamefont
  {Hikihara}, \citenamefont {Kecke}, \citenamefont {Momoi},\ and\ \citenamefont
  {Furusaki}}]{j1j2_H2}%
  \BibitemOpen
  \bibfield  {author} {\bibinfo {author} {\bibfnamefont {T.}~\bibnamefont
  {Hikihara}}, \bibinfo {author} {\bibfnamefont {L.}~\bibnamefont {Kecke}},
  \bibinfo {author} {\bibfnamefont {T.}~\bibnamefont {Momoi}}, \ and\ \bibinfo
  {author} {\bibfnamefont {A.}~\bibnamefont {Furusaki}},\ }\href {\doibase
  10.1103/PhysRevB.78.144404} {\bibfield  {journal} {\bibinfo  {journal} {Phys.
  Rev. B}\ }\textbf {\bibinfo {volume} {78}},\ \bibinfo {pages} {144404}
  (\bibinfo {year} {2008})}\BibitemShut {NoStop}%
\bibitem [{\citenamefont {Zhitomirsky}\ and\ \citenamefont
  {Tsunetsugu}(2010)}]{zhitomirsky2010}%
  \BibitemOpen
  \bibfield  {author} {\bibinfo {author} {\bibfnamefont {M.~E.}\ \bibnamefont
  {Zhitomirsky}}\ and\ \bibinfo {author} {\bibfnamefont {H.}~\bibnamefont
  {Tsunetsugu}},\ }\href@noop {} {\bibfield  {journal} {\bibinfo  {journal}
  {Europhys. Lett.}\ }\textbf {\bibinfo {volume} {92}},\ \bibinfo {pages}
  {37001} (\bibinfo {year} {2010})}\BibitemShut {NoStop}%
\bibitem [{\citenamefont {Sato}\ \emph {et~al.}(2013)\citenamefont {Sato},
  \citenamefont {Hikihara},\ and\ \citenamefont {Momoi}}]{sato2013}%
  \BibitemOpen
  \bibfield  {author} {\bibinfo {author} {\bibfnamefont {M.}~\bibnamefont
  {Sato}}, \bibinfo {author} {\bibfnamefont {T.}~\bibnamefont {Hikihara}}, \
  and\ \bibinfo {author} {\bibfnamefont {T.}~\bibnamefont {Momoi}},\
  }\href@noop {} {\bibfield  {journal} {\bibinfo  {journal} {Phys. Rev. Lett.}\
  }\textbf {\bibinfo {volume} {110}},\ \bibinfo {pages} {077206} (\bibinfo
  {year} {2013})}\BibitemShut {NoStop}%
\bibitem [{\citenamefont {Willenberg}\ \emph {et~al.}(2016)\citenamefont
  {Willenberg}, \citenamefont {Sch\"apers}, \citenamefont {Wolter},
  \citenamefont {Drechsler}, \citenamefont {Reehuis}, \citenamefont {Hoffmann},
  \citenamefont {B\"uchner}, \citenamefont {Studer}, \citenamefont {Rule},
  \citenamefont {Ouladdiaf}, \citenamefont {S\"ullow},\ and\ \citenamefont
  {Nishimoto}}]{willenberg2016}%
  \BibitemOpen
  \bibfield  {author} {\bibinfo {author} {\bibfnamefont {B.}~\bibnamefont
  {Willenberg}}, \bibinfo {author} {\bibfnamefont {M.}~\bibnamefont
  {Sch\"apers}}, \bibinfo {author} {\bibfnamefont {A.~U.~B.}\ \bibnamefont
  {Wolter}}, \bibinfo {author} {\bibfnamefont {S.-L.}\ \bibnamefont
  {Drechsler}}, \bibinfo {author} {\bibfnamefont {M.}~\bibnamefont {Reehuis}},
  \bibinfo {author} {\bibfnamefont {J.-U.}\ \bibnamefont {Hoffmann}}, \bibinfo
  {author} {\bibfnamefont {B.}~\bibnamefont {B\"uchner}}, \bibinfo {author}
  {\bibfnamefont {A.~J.}\ \bibnamefont {Studer}}, \bibinfo {author}
  {\bibfnamefont {K.~C.}\ \bibnamefont {Rule}}, \bibinfo {author}
  {\bibfnamefont {B.}~\bibnamefont {Ouladdiaf}}, \bibinfo {author}
  {\bibfnamefont {S.}~\bibnamefont {S\"ullow}}, \ and\ \bibinfo {author}
  {\bibfnamefont {S.}~\bibnamefont {Nishimoto}},\ }\href@noop {} {\bibfield
  {journal} {\bibinfo  {journal} {Phys. Rev. Lett.}\ }\textbf {\bibinfo
  {volume} {116}},\ \bibinfo {pages} {047202} (\bibinfo {year}
  {2016})}\BibitemShut {NoStop}%
\bibitem [{\citenamefont {Weickert}\ \emph {et~al.}()\citenamefont {Weickert},
  \citenamefont {Harrison}, \citenamefont {Scott}, \citenamefont {Jaime},
  \citenamefont {Leitm\"ae}, \citenamefont {Heinmaa}, \citenamefont {Stern},
  \citenamefont {Janson}, \citenamefont {Berger}, \citenamefont {Rosner},\ and\
  \citenamefont {Tsirlin}}]{weickert2016}%
  \BibitemOpen
  \bibfield  {author} {\bibinfo {author} {\bibfnamefont {F.}~\bibnamefont
  {Weickert}}, \bibinfo {author} {\bibfnamefont {N.}~\bibnamefont {Harrison}},
  \bibinfo {author} {\bibfnamefont {B.}~\bibnamefont {Scott}}, \bibinfo
  {author} {\bibfnamefont {M.}~\bibnamefont {Jaime}}, \bibinfo {author}
  {\bibfnamefont {A.}~\bibnamefont {Leitm\"ae}}, \bibinfo {author}
  {\bibfnamefont {I.}~\bibnamefont {Heinmaa}}, \bibinfo {author} {\bibfnamefont
  {R.}~\bibnamefont {Stern}}, \bibinfo {author} {\bibfnamefont
  {O.}~\bibnamefont {Janson}}, \bibinfo {author} {\bibfnamefont
  {H.}~\bibnamefont {Berger}}, \bibinfo {author} {\bibfnamefont
  {H.}~\bibnamefont {Rosner}}, \ and\ \bibinfo {author} {\bibfnamefont
  {A.}~\bibnamefont {Tsirlin}},\ }\href@noop {} {\enquote {\bibinfo {title}
  {Magnetic anisotropy in the frustrated spin chain compound
  {$\beta$-TeVO$_4$}},}\ }\bibinfo {note} {{a}rXiv:1602.01632}\BibitemShut
  {NoStop}%
\bibitem [{\citenamefont {Eggert}(1996)}]{eggert1996}%
  \BibitemOpen
  \bibfield  {author} {\bibinfo {author} {\bibfnamefont {S.}~\bibnamefont
  {Eggert}},\ }\href@noop {} {\bibfield  {journal} {\bibinfo  {journal} {Phys.
  Rev. B}\ }\textbf {\bibinfo {volume} {54}},\ \bibinfo {pages} {R9612}
  (\bibinfo {year} {1996})}\BibitemShut {NoStop}%
\bibitem [{\citenamefont {White}\ and\ \citenamefont
  {Affleck}(1996)}]{white1996}%
  \BibitemOpen
  \bibfield  {author} {\bibinfo {author} {\bibfnamefont {S.~R.}\ \bibnamefont
  {White}}\ and\ \bibinfo {author} {\bibfnamefont {I.}~\bibnamefont
  {Affleck}},\ }\href@noop {} {\bibfield  {journal} {\bibinfo  {journal} {Phys.
  Rev. B}\ }\textbf {\bibinfo {volume} {54}},\ \bibinfo {pages} {9862}
  (\bibinfo {year} {1996})},\ \bibinfo {note} {cond-mat/9602126}\BibitemShut
  {NoStop}%
\bibitem [{\citenamefont {Majumdar}\ and\ \citenamefont
  {Ghosh}(1969{\natexlab{a}})}]{majumdar1969}%
  \BibitemOpen
  \bibfield  {author} {\bibinfo {author} {\bibfnamefont {C.~K.}\ \bibnamefont
  {Majumdar}}\ and\ \bibinfo {author} {\bibfnamefont {D.~K.}\ \bibnamefont
  {Ghosh}},\ }\href@noop {} {\bibfield  {journal} {\bibinfo  {journal} {J.
  Math. Phys.}\ }\textbf {\bibinfo {volume} {10}},\ \bibinfo {pages} {1388}
  (\bibinfo {year} {1969}{\natexlab{a}})}\BibitemShut {NoStop}%
\bibitem [{\citenamefont {Majumdar}\ and\ \citenamefont
  {Ghosh}(1969{\natexlab{b}})}]{majumdar1969b}%
  \BibitemOpen
  \bibfield  {author} {\bibinfo {author} {\bibfnamefont {C.~K.}\ \bibnamefont
  {Majumdar}}\ and\ \bibinfo {author} {\bibfnamefont {D.~K.}\ \bibnamefont
  {Ghosh}},\ }\href@noop {} {\bibfield  {journal} {\bibinfo  {journal} {J.
  Math. Phys.}\ }\textbf {\bibinfo {volume} {10}},\ \bibinfo {pages} {1399}
  (\bibinfo {year} {1969}{\natexlab{b}})}\BibitemShut {NoStop}%
\bibitem [{\citenamefont {Hase}\ \emph {et~al.}(1993)\citenamefont {Hase},
  \citenamefont {Terasaki},\ and\ \citenamefont {Uchinokura}}]{hase1993}%
  \BibitemOpen
  \bibfield  {author} {\bibinfo {author} {\bibfnamefont {M.}~\bibnamefont
  {Hase}}, \bibinfo {author} {\bibfnamefont {I.}~\bibnamefont {Terasaki}}, \
  and\ \bibinfo {author} {\bibfnamefont {K.}~\bibnamefont {Uchinokura}},\
  }\href@noop {} {\bibfield  {journal} {\bibinfo  {journal} {Phys. Rev. Lett}\
  }\textbf {\bibinfo {volume} {70}},\ \bibinfo {pages} {3651} (\bibinfo {year}
  {1993})}\BibitemShut {NoStop}%
\bibitem [{\citenamefont {Kiryukhin}\ \emph {et~al.}(1996)\citenamefont
  {Kiryukhin}, \citenamefont {Keimer}, \citenamefont {Hill},\ and\
  \citenamefont {Vigliante}}]{kiryukhin1996}%
  \BibitemOpen
  \bibfield  {author} {\bibinfo {author} {\bibfnamefont {V.}~\bibnamefont
  {Kiryukhin}}, \bibinfo {author} {\bibfnamefont {B.}~\bibnamefont {Keimer}},
  \bibinfo {author} {\bibfnamefont {J.~P.}\ \bibnamefont {Hill}}, \ and\
  \bibinfo {author} {\bibfnamefont {A.}~\bibnamefont {Vigliante}},\ }\href@noop
  {} {\bibfield  {journal} {\bibinfo  {journal} {Phys. Rev. Lett.}\ }\textbf
  {\bibinfo {volume} {76}},\ \bibinfo {pages} {4608} (\bibinfo {year}
  {1996})}\BibitemShut {NoStop}%
\bibitem [{\citenamefont {Brown}\ \emph {et~al.}(1979)\citenamefont {Brown},
  \citenamefont {Donner}, \citenamefont {Hall}, \citenamefont {Wilson},
  \citenamefont {Wilson}, \citenamefont {Hodgson},\ and\ \citenamefont
  {Hatfield}}]{brown1979}%
  \BibitemOpen
  \bibfield  {author} {\bibinfo {author} {\bibfnamefont {D.~B.}\ \bibnamefont
  {Brown}}, \bibinfo {author} {\bibfnamefont {J.~A.}\ \bibnamefont {Donner}},
  \bibinfo {author} {\bibfnamefont {J.~W.}\ \bibnamefont {Hall}}, \bibinfo
  {author} {\bibfnamefont {S.~R.}\ \bibnamefont {Wilson}}, \bibinfo {author}
  {\bibfnamefont {R.~B.}\ \bibnamefont {Wilson}}, \bibinfo {author}
  {\bibfnamefont {D.~J.}\ \bibnamefont {Hodgson}}, \ and\ \bibinfo {author}
  {\bibfnamefont {W.~E.}\ \bibnamefont {Hatfield}},\ }\href@noop {} {\bibfield
  {journal} {\bibinfo  {journal} {Inorg. Chem.}\ }\textbf {\bibinfo {volume}
  {18}},\ \bibinfo {pages} {2635} (\bibinfo {year} {1979})}\BibitemShut
  {NoStop}%
\bibitem [{\citenamefont {Hagiwara}\ \emph {et~al.}(2001)\citenamefont
  {Hagiwara}, \citenamefont {Narumi}, \citenamefont {Kindo}, \citenamefont
  {Maeshima}, \citenamefont {Okunishi}, \citenamefont {Sakai},\ and\
  \citenamefont {Takahashi}}]{hagiwara2001}%
  \BibitemOpen
  \bibfield  {author} {\bibinfo {author} {\bibfnamefont {M.}~\bibnamefont
  {Hagiwara}}, \bibinfo {author} {\bibfnamefont {Y.}~\bibnamefont {Narumi}},
  \bibinfo {author} {\bibfnamefont {K.}~\bibnamefont {Kindo}}, \bibinfo
  {author} {\bibfnamefont {N.}~\bibnamefont {Maeshima}}, \bibinfo {author}
  {\bibfnamefont {K.}~\bibnamefont {Okunishi}}, \bibinfo {author}
  {\bibfnamefont {T.}~\bibnamefont {Sakai}}, \ and\ \bibinfo {author}
  {\bibfnamefont {M.}~\bibnamefont {Takahashi}},\ }\href@noop {} {\bibfield
  {journal} {\bibinfo  {journal} {Physica B}\ }\textbf {\bibinfo {volume}
  {294--295}},\ \bibinfo {pages} {83} (\bibinfo {year} {2001})}\BibitemShut
  {NoStop}%
\bibitem [{\citenamefont {Maeshima}\ \emph {et~al.}(2003)\citenamefont
  {Maeshima}, \citenamefont {Hagiwara}, \citenamefont {Narumi}, \citenamefont
  {Kindo}, \citenamefont {Kobayashi},\ and\ \citenamefont
  {Okunishi}}]{maeshima2003}%
  \BibitemOpen
  \bibfield  {author} {\bibinfo {author} {\bibfnamefont {N.}~\bibnamefont
  {Maeshima}}, \bibinfo {author} {\bibfnamefont {M.}~\bibnamefont {Hagiwara}},
  \bibinfo {author} {\bibfnamefont {Y.}~\bibnamefont {Narumi}}, \bibinfo
  {author} {\bibfnamefont {K.}~\bibnamefont {Kindo}}, \bibinfo {author}
  {\bibfnamefont {T.~C.}\ \bibnamefont {Kobayashi}}, \ and\ \bibinfo {author}
  {\bibfnamefont {K.}~\bibnamefont {Okunishi}},\ }\href@noop {} {\bibfield
  {journal} {\bibinfo  {journal} {J. Phys.: Condens. Matter}\ }\textbf
  {\bibinfo {volume} {15}},\ \bibinfo {pages} {3607} (\bibinfo {year}
  {2003})}\BibitemShut {NoStop}%
\bibitem [{\citenamefont {Nilsen}\ \emph {et~al.}(2008)\citenamefont {Nilsen},
  \citenamefont {R{\o}nnow}, \citenamefont {L{\"a}uchli}, \citenamefont
  {Fabbiani}, \citenamefont {Sanchez-Benitez}, \citenamefont {Kamenev},\ and\
  \citenamefont {Harrison}}]{nilsen2008}%
  \BibitemOpen
  \bibfield  {author} {\bibinfo {author} {\bibfnamefont {G.~J.}\ \bibnamefont
  {Nilsen}}, \bibinfo {author} {\bibfnamefont {H.~M.}\ \bibnamefont
  {R{\o}nnow}}, \bibinfo {author} {\bibfnamefont {A.~M.}\ \bibnamefont
  {L{\"a}uchli}}, \bibinfo {author} {\bibfnamefont {F.~P.~A.}\ \bibnamefont
  {Fabbiani}}, \bibinfo {author} {\bibfnamefont {J.}~\bibnamefont
  {Sanchez-Benitez}}, \bibinfo {author} {\bibfnamefont {K.~V.}\ \bibnamefont
  {Kamenev}}, \ and\ \bibinfo {author} {\bibfnamefont {A.}~\bibnamefont
  {Harrison}},\ }\href@noop {} {\bibfield  {journal} {\bibinfo  {journal}
  {Chem. Mater.}\ }\textbf {\bibinfo {volume} {20}},\ \bibinfo {pages} {8}
  (\bibinfo {year} {2008})}\BibitemShut {NoStop}%
\bibitem [{\citenamefont {Kasinathan}\ \emph {et~al.}(2013)\citenamefont
  {Kasinathan}, \citenamefont {Koepernik}, \citenamefont {Janson},
  \citenamefont {Nilsen}, \citenamefont {Piatek}, \citenamefont {R{\o}nnow},\
  and\ \citenamefont {Rosner}}]{kasinathan2013}%
  \BibitemOpen
  \bibfield  {author} {\bibinfo {author} {\bibfnamefont {D.}~\bibnamefont
  {Kasinathan}}, \bibinfo {author} {\bibfnamefont {K.}~\bibnamefont
  {Koepernik}}, \bibinfo {author} {\bibfnamefont {O.}~\bibnamefont {Janson}},
  \bibinfo {author} {\bibfnamefont {G.~J.}\ \bibnamefont {Nilsen}}, \bibinfo
  {author} {\bibfnamefont {J.~O.}\ \bibnamefont {Piatek}}, \bibinfo {author}
  {\bibfnamefont {H.~M.}\ \bibnamefont {R{\o}nnow}}, \ and\ \bibinfo {author}
  {\bibfnamefont {H.}~\bibnamefont {Rosner}},\ }\href@noop {} {\bibfield
  {journal} {\bibinfo  {journal} {Phys. Rev. B}\ }\textbf {\bibinfo {volume}
  {88}},\ \bibinfo {pages} {224410} (\bibinfo {year} {2013})}\BibitemShut
  {NoStop}%
\bibitem [{\citenamefont {Fujisawa}\ \emph {et~al.}(2011)\citenamefont
  {Fujisawa}, \citenamefont {Kikuchi}, \citenamefont {Fujii}, \citenamefont
  {Mitsudo}, \citenamefont {Matsuo},\ and\ \citenamefont {Kindo}}]{szen_1}%
  \BibitemOpen
  \bibfield  {author} {\bibinfo {author} {\bibfnamefont {M.}~\bibnamefont
  {Fujisawa}}, \bibinfo {author} {\bibfnamefont {H.}~\bibnamefont {Kikuchi}},
  \bibinfo {author} {\bibfnamefont {Y.}~\bibnamefont {Fujii}}, \bibinfo
  {author} {\bibfnamefont {S.}~\bibnamefont {Mitsudo}}, \bibinfo {author}
  {\bibfnamefont {A.}~\bibnamefont {Matsuo}}, \ and\ \bibinfo {author}
  {\bibfnamefont {K.}~\bibnamefont {Kindo}},\ }\href@noop {} {\bibfield
  {journal} {\bibinfo  {journal} {J. Phys.: Conf. Series}\ }\textbf {\bibinfo
  {volume} {320}},\ \bibinfo {pages} {012031} (\bibinfo {year}
  {2011})}\BibitemShut {NoStop}%
\bibitem [{\citenamefont {Fujii}\ \emph {et~al.}(2015)\citenamefont {Fujii},
  \citenamefont {Kikuchi}, \citenamefont {Nakagawa}, \citenamefont {Takada},\
  and\ \citenamefont {Fujisawa}}]{nmr_szen}%
  \BibitemOpen
  \bibfield  {author} {\bibinfo {author} {\bibfnamefont {Y.}~\bibnamefont
  {Fujii}}, \bibinfo {author} {\bibfnamefont {H.}~\bibnamefont {Kikuchi}},
  \bibinfo {author} {\bibfnamefont {K.}~\bibnamefont {Nakagawa}}, \bibinfo
  {author} {\bibfnamefont {S.}~\bibnamefont {Takada}}, \ and\ \bibinfo {author}
  {\bibfnamefont {M.}~\bibnamefont {Fujisawa}},\ }\href@noop {} {\bibfield
  {journal} {\bibinfo  {journal} {Physics Procedia}\ }\textbf {\bibinfo
  {volume} {75}},\ \bibinfo {pages} {589–} (\bibinfo {year}
  {2015})}\BibitemShut {NoStop}%
\bibitem [{\citenamefont {Vilminot}\ \emph {et~al.}(2003)\citenamefont
  {Vilminot}, \citenamefont {Richard-Plouet}, \citenamefont {Andr\'e},
  \citenamefont {Swierczynski}, \citenamefont {Guillot}, \citenamefont
  {Bour\'ee-Vigneron},\ and\ \citenamefont {Drillon}}]{antlerite_2003}%
  \BibitemOpen
  \bibfield  {author} {\bibinfo {author} {\bibfnamefont {S.}~\bibnamefont
  {Vilminot}}, \bibinfo {author} {\bibfnamefont {M.}~\bibnamefont
  {Richard-Plouet}}, \bibinfo {author} {\bibfnamefont {G.}~\bibnamefont
  {Andr\'e}}, \bibinfo {author} {\bibfnamefont {D.}~\bibnamefont
  {Swierczynski}}, \bibinfo {author} {\bibfnamefont {M.}~\bibnamefont
  {Guillot}}, \bibinfo {author} {\bibfnamefont {F.}~\bibnamefont
  {Bour\'ee-Vigneron}}, \ and\ \bibinfo {author} {\bibfnamefont
  {M.}~\bibnamefont {Drillon}},\ }\href@noop {} {\bibfield  {journal} {\bibinfo
   {journal} {J. Solid State Chem.}\ }\textbf {\bibinfo {volume} {170}},\
  \bibinfo {pages} {255 } (\bibinfo {year} {2003})}\BibitemShut {NoStop}%
\bibitem [{\citenamefont {Vilminot}\ \emph {et~al.}(2007)\citenamefont
  {Vilminot}, \citenamefont {Andr\'e}, \citenamefont {Bour\'ee-Vigneron},
  \citenamefont {Richard-Plouet},\ and\ \citenamefont
  {Kurmoo}}]{antlerite_2007}%
  \BibitemOpen
  \bibfield  {author} {\bibinfo {author} {\bibfnamefont {S.}~\bibnamefont
  {Vilminot}}, \bibinfo {author} {\bibfnamefont {G.}~\bibnamefont {Andr\'e}},
  \bibinfo {author} {\bibfnamefont {F.}~\bibnamefont {Bour\'ee-Vigneron}},
  \bibinfo {author} {\bibfnamefont {M.}~\bibnamefont {Richard-Plouet}}, \ and\
  \bibinfo {author} {\bibfnamefont {M.}~\bibnamefont {Kurmoo}},\ }\href@noop {}
  {\bibfield  {journal} {\bibinfo  {journal} {Inorg. Chem.}\ }\textbf {\bibinfo
  {volume} {46}},\ \bibinfo {pages} {10079} (\bibinfo {year}
  {2007})}\BibitemShut {NoStop}%
\bibitem [{sup()}]{supplement}%
  \BibitemOpen
  \href@noop {} {}\bibinfo {note} {See supplementary material for the XRD
  powder pattern, high resolution crystallographic data from synchrotron
  XRD-diffraction, $\chi(T)$ in different magnetic fields, the back ground
  contributions to $C_p(T)$ at H=0\,T}\BibitemShut {NoStop}%
\bibitem [{\citenamefont {Stolz}\ and\ \citenamefont
  {Armbruster}(1998)}]{xxstr}%
  \BibitemOpen
  \bibfield  {author} {\bibinfo {author} {\bibfnamefont {J.}~\bibnamefont
  {Stolz}}\ and\ \bibinfo {author} {\bibfnamefont {T.}~\bibnamefont
  {Armbruster}},\ }\href@noop {} {\bibfield  {journal} {\bibinfo  {journal}
  {Neues Jahrb. Mineral. (Monatsh.)}\ ,\ \bibinfo {pages} {278}} (\bibinfo
  {year} {1998})}\BibitemShut {NoStop}%
\bibitem [{\citenamefont {Koepernik}\ and\ \citenamefont
  {Eschrig}(1999)}]{fplo}%
  \BibitemOpen
  \bibfield  {author} {\bibinfo {author} {\bibfnamefont {K.}~\bibnamefont
  {Koepernik}}\ and\ \bibinfo {author} {\bibfnamefont {H.}~\bibnamefont
  {Eschrig}},\ }\href@noop {} {\bibfield  {journal} {\bibinfo  {journal} {Phys.
  Rev. B}\ }\textbf {\bibinfo {volume} {59}},\ \bibinfo {pages} {1743}
  (\bibinfo {year} {1999})}\BibitemShut {NoStop}%
\bibitem [{\citenamefont {Perdew}\ and\ \citenamefont {Wang}(1992)}]{pw92}%
  \BibitemOpen
  \bibfield  {author} {\bibinfo {author} {\bibfnamefont {J.~P.}\ \bibnamefont
  {Perdew}}\ and\ \bibinfo {author} {\bibfnamefont {Y.}~\bibnamefont {Wang}},\
  }\href@noop {} {\bibfield  {journal} {\bibinfo  {journal} {Phys. Rev. B}\
  }\textbf {\bibinfo {volume} {45}},\ \bibinfo {pages} {13244} (\bibinfo {year}
  {1992})}\BibitemShut {NoStop}%
\bibitem [{\citenamefont {Perdew}\ \emph {et~al.}(1996)\citenamefont {Perdew},
  \citenamefont {Burke},\ and\ \citenamefont {Ernzerhof}}]{pbe96}%
  \BibitemOpen
  \bibfield  {author} {\bibinfo {author} {\bibfnamefont {J.~P.}\ \bibnamefont
  {Perdew}}, \bibinfo {author} {\bibfnamefont {K.}~\bibnamefont {Burke}}, \
  and\ \bibinfo {author} {\bibfnamefont {M.}~\bibnamefont {Ernzerhof}},\
  }\href@noop {} {\bibfield  {journal} {\bibinfo  {journal} {Phys. Rev. Lett.}\
  }\textbf {\bibinfo {volume} {77}},\ \bibinfo {pages} {3865} (\bibinfo {year}
  {1996})}\BibitemShut {NoStop}%
\bibitem [{\citenamefont {Anisimov}\ \emph {et~al.}(1991)\citenamefont
  {Anisimov}, \citenamefont {Zaanen},\ and\ \citenamefont {Andersen}}]{lsdau1}%
  \BibitemOpen
  \bibfield  {author} {\bibinfo {author} {\bibfnamefont {V.~I.}\ \bibnamefont
  {Anisimov}}, \bibinfo {author} {\bibfnamefont {J.}~\bibnamefont {Zaanen}}, \
  and\ \bibinfo {author} {\bibfnamefont {O.~K.}\ \bibnamefont {Andersen}},\
  }\href {\doibase 10.1103/PhysRevB.44.943} {\bibfield  {journal} {\bibinfo
  {journal} {Phys. Rev. B}\ }\textbf {\bibinfo {volume} {44}},\ \bibinfo
  {pages} {943} (\bibinfo {year} {1991})}\BibitemShut {NoStop}%
\bibitem [{\citenamefont {Anisimov}\ \emph {et~al.}(1997)\citenamefont
  {Anisimov}, \citenamefont {Aryasetiawan},\ and\ \citenamefont
  {Lichtenstein}}]{lsdau2}%
  \BibitemOpen
  \bibfield  {author} {\bibinfo {author} {\bibfnamefont {V.~I.}\ \bibnamefont
  {Anisimov}}, \bibinfo {author} {\bibfnamefont {F.}~\bibnamefont
  {Aryasetiawan}}, \ and\ \bibinfo {author} {\bibfnamefont {A.~I.}\
  \bibnamefont {Lichtenstein}},\ }\href@noop {} {\bibfield  {journal} {\bibinfo
   {journal} {J. Phys.: Condens. Matter}\ }\textbf {\bibinfo {volume} {9}},\
  \bibinfo {pages} {767} (\bibinfo {year} {1997})}\BibitemShut {NoStop}%
\bibitem [{\citenamefont {Lebernegg}\ \emph
  {et~al.}(2013{\natexlab{a}})\citenamefont {Lebernegg}, \citenamefont
  {Tsirlin}, \citenamefont {Janson},\ and\ \citenamefont {Rosner}}]{malachite}%
  \BibitemOpen
  \bibfield  {author} {\bibinfo {author} {\bibfnamefont {S.}~\bibnamefont
  {Lebernegg}}, \bibinfo {author} {\bibfnamefont {A.~A.}\ \bibnamefont
  {Tsirlin}}, \bibinfo {author} {\bibfnamefont {O.}~\bibnamefont {Janson}}, \
  and\ \bibinfo {author} {\bibfnamefont {H.}~\bibnamefont {Rosner}},\ }\href
  {\doibase 10.1103/PhysRevB.88.224406} {\bibfield  {journal} {\bibinfo
  {journal} {Phys. Rev. B}\ }\textbf {\bibinfo {volume} {88}},\ \bibinfo
  {pages} {224406} (\bibinfo {year} {2013}{\natexlab{a}})}\BibitemShut
  {NoStop}%
\bibitem [{\citenamefont {Lebernegg}\ \emph
  {et~al.}(2013{\natexlab{b}})\citenamefont {Lebernegg}, \citenamefont
  {Tsirlin}, \citenamefont {Janson},\ and\ \citenamefont
  {Rosner}}]{clinoclase}%
  \BibitemOpen
  \bibfield  {author} {\bibinfo {author} {\bibfnamefont {S.}~\bibnamefont
  {Lebernegg}}, \bibinfo {author} {\bibfnamefont {A.~A.}\ \bibnamefont
  {Tsirlin}}, \bibinfo {author} {\bibfnamefont {O.}~\bibnamefont {Janson}}, \
  and\ \bibinfo {author} {\bibfnamefont {H.}~\bibnamefont {Rosner}},\ }\href
  {\doibase 10.1103/PhysRevB.87.235117} {\bibfield  {journal} {\bibinfo
  {journal} {Phys. Rev. B}\ }\textbf {\bibinfo {volume} {87}},\ \bibinfo
  {pages} {235117} (\bibinfo {year} {2013}{\natexlab{b}})}\BibitemShut
  {NoStop}%
\bibitem [{\citenamefont {Tsirlin}\ \emph {et~al.}(2013)\citenamefont
  {Tsirlin}, \citenamefont {Janson}, \citenamefont {Lebernegg},\ and\
  \citenamefont {Rosner}}]{diaboleite}%
  \BibitemOpen
  \bibfield  {author} {\bibinfo {author} {\bibfnamefont {A.~A.}\ \bibnamefont
  {Tsirlin}}, \bibinfo {author} {\bibfnamefont {O.}~\bibnamefont {Janson}},
  \bibinfo {author} {\bibfnamefont {S.}~\bibnamefont {Lebernegg}}, \ and\
  \bibinfo {author} {\bibfnamefont {H.}~\bibnamefont {Rosner}},\ }\href
  {\doibase 10.1103/PhysRevB.87.064404} {\bibfield  {journal} {\bibinfo
  {journal} {Phys. Rev. B}\ }\textbf {\bibinfo {volume} {87}},\ \bibinfo
  {pages} {064404} (\bibinfo {year} {2013})}\BibitemShut {NoStop}%
\bibitem [{\citenamefont {Illas}\ \emph {et~al.}(2000)\citenamefont {Illas},
  \citenamefont {Moreira}, \citenamefont {{de Graaf}},\ and\ \citenamefont
  {Barone}}]{bs2}%
  \BibitemOpen
  \bibfield  {author} {\bibinfo {author} {\bibfnamefont {F.}~\bibnamefont
  {Illas}}, \bibinfo {author} {\bibfnamefont {I.}~\bibnamefont {Moreira}},
  \bibinfo {author} {\bibfnamefont {C.}~\bibnamefont {{de Graaf}}}, \ and\
  \bibinfo {author} {\bibfnamefont {V.}~\bibnamefont {Barone}},\ }\href@noop {}
  {\bibfield  {journal} {\bibinfo  {journal} {Theor. Chem. Acc.}\ }\textbf
  {\bibinfo {volume} {104}},\ \bibinfo {pages} {265} (\bibinfo {year}
  {2000})}\BibitemShut {NoStop}%
\bibitem [{\citenamefont {Janson}\ \emph {et~al.}(2010)\citenamefont {Janson},
  \citenamefont {Tsirlin}, \citenamefont {Schmitt},\ and\ \citenamefont
  {Rosner}}]{dioptase}%
  \BibitemOpen
  \bibfield  {author} {\bibinfo {author} {\bibfnamefont {O.}~\bibnamefont
  {Janson}}, \bibinfo {author} {\bibfnamefont {A.~A.}\ \bibnamefont {Tsirlin}},
  \bibinfo {author} {\bibfnamefont {M.}~\bibnamefont {Schmitt}}, \ and\
  \bibinfo {author} {\bibfnamefont {H.}~\bibnamefont {Rosner}},\ }\href
  {\doibase 10.1103/PhysRevB.82.014424} {\bibfield  {journal} {\bibinfo
  {journal} {Phys. Rev. B}\ }\textbf {\bibinfo {volume} {82}},\ \bibinfo
  {pages} {014424} (\bibinfo {year} {2010})}\BibitemShut {NoStop}%
\bibitem [{\citenamefont {Lebernegg}\ \emph {et~al.}(2014)\citenamefont
  {Lebernegg}, \citenamefont {Tsirlin}, \citenamefont {Janson},\ and\
  \citenamefont {Rosner}}]{callaghanite}%
  \BibitemOpen
  \bibfield  {author} {\bibinfo {author} {\bibfnamefont {S.}~\bibnamefont
  {Lebernegg}}, \bibinfo {author} {\bibfnamefont {A.~A.}\ \bibnamefont
  {Tsirlin}}, \bibinfo {author} {\bibfnamefont {O.}~\bibnamefont {Janson}}, \
  and\ \bibinfo {author} {\bibfnamefont {H.}~\bibnamefont {Rosner}},\ }\href
  {\doibase 10.1103/PhysRevB.89.165127} {\bibfield  {journal} {\bibinfo
  {journal} {Phys. Rev. B}\ }\textbf {\bibinfo {volume} {89}},\ \bibinfo
  {pages} {165127} (\bibinfo {year} {2014})}\BibitemShut {NoStop}%
\bibitem [{\citenamefont {Albuquerque}\ \emph {et~al.}(2007)\citenamefont
  {Albuquerque}, \citenamefont {Alet}, \citenamefont {Corboz}, \citenamefont
  {Dayal}, \citenamefont {Feiguin}, \citenamefont {Fuchs}, \citenamefont
  {Gamper}, \citenamefont {Gull}, \citenamefont {G\"urtler}, \citenamefont
  {Honecker}, \citenamefont {Igarashi}, \citenamefont {K\"orner}, \citenamefont
  {Kozhevnikov}, \citenamefont {L\"auchli}, \citenamefont {Manmana},
  \citenamefont {Matsumoto}, \citenamefont {McCulloch}, \citenamefont {Michel},
  \citenamefont {Noack}, \citenamefont {Paw{\l}owski}, \citenamefont {Pollet},
  \citenamefont {Pruschke}, \citenamefont {Schollw\"ock}, \citenamefont {Todo},
  \citenamefont {Trebst}, \citenamefont {Troyer}, \citenamefont {Werner},\ and\
  \citenamefont {Wessel}}]{ALPS}%
  \BibitemOpen
  \bibfield  {author} {\bibinfo {author} {\bibfnamefont {A.}~\bibnamefont
  {Albuquerque}}, \bibinfo {author} {\bibfnamefont {F.}~\bibnamefont {Alet}},
  \bibinfo {author} {\bibfnamefont {P.}~\bibnamefont {Corboz}}, \bibinfo
  {author} {\bibfnamefont {P.}~\bibnamefont {Dayal}}, \bibinfo {author}
  {\bibfnamefont {A.}~\bibnamefont {Feiguin}}, \bibinfo {author} {\bibfnamefont
  {S.}~\bibnamefont {Fuchs}}, \bibinfo {author} {\bibfnamefont
  {L.}~\bibnamefont {Gamper}}, \bibinfo {author} {\bibfnamefont
  {E.}~\bibnamefont {Gull}}, \bibinfo {author} {\bibfnamefont {S.}~\bibnamefont
  {G\"urtler}}, \bibinfo {author} {\bibfnamefont {A.}~\bibnamefont {Honecker}},
  \bibinfo {author} {\bibfnamefont {R.}~\bibnamefont {Igarashi}}, \bibinfo
  {author} {\bibfnamefont {M.}~\bibnamefont {K\"orner}}, \bibinfo {author}
  {\bibfnamefont {A.}~\bibnamefont {Kozhevnikov}}, \bibinfo {author}
  {\bibfnamefont {A.}~\bibnamefont {L\"auchli}}, \bibinfo {author}
  {\bibfnamefont {S.~R.}\ \bibnamefont {Manmana}}, \bibinfo {author}
  {\bibfnamefont {M.}~\bibnamefont {Matsumoto}}, \bibinfo {author}
  {\bibfnamefont {I.~P.}\ \bibnamefont {McCulloch}}, \bibinfo {author}
  {\bibfnamefont {F.}~\bibnamefont {Michel}}, \bibinfo {author} {\bibfnamefont
  {R.~M.}\ \bibnamefont {Noack}}, \bibinfo {author} {\bibfnamefont
  {G.}~\bibnamefont {Paw{\l}owski}}, \bibinfo {author} {\bibfnamefont
  {L.}~\bibnamefont {Pollet}}, \bibinfo {author} {\bibfnamefont
  {T.}~\bibnamefont {Pruschke}}, \bibinfo {author} {\bibfnamefont
  {U.}~\bibnamefont {Schollw\"ock}}, \bibinfo {author} {\bibfnamefont
  {S.}~\bibnamefont {Todo}}, \bibinfo {author} {\bibfnamefont {S.}~\bibnamefont
  {Trebst}}, \bibinfo {author} {\bibfnamefont {M.}~\bibnamefont {Troyer}},
  \bibinfo {author} {\bibfnamefont {P.}~\bibnamefont {Werner}}, \ and\ \bibinfo
  {author} {\bibfnamefont {S.}~\bibnamefont {Wessel}},\ }\href {\doibase
  10.1016/j.jmmm.2006.10.304} {\bibfield  {journal} {\bibinfo  {journal} {J.
  Magn. Magn. Mater.}\ }\textbf {\bibinfo {volume} {310}},\ \bibinfo {pages}
  {1187} (\bibinfo {year} {2007})}\BibitemShut {NoStop}%
\bibitem [{\citenamefont {Wang}\ and\ \citenamefont {Xiang}(1997)}]{wang1997}%
  \BibitemOpen
  \bibfield  {author} {\bibinfo {author} {\bibfnamefont {X.}~\bibnamefont
  {Wang}}\ and\ \bibinfo {author} {\bibfnamefont {T.}~\bibnamefont {Xiang}},\
  }\href@noop {} {\bibfield  {journal} {\bibinfo  {journal} {Phys. Rev. B}\
  }\textbf {\bibinfo {volume} {56}},\ \bibinfo {pages} {5061} (\bibinfo {year}
  {1997})}\BibitemShut {NoStop}%
\bibitem [{\citenamefont {Shibata}(1997)}]{shibata1997}%
  \BibitemOpen
  \bibfield  {author} {\bibinfo {author} {\bibfnamefont {N.}~\bibnamefont
  {Shibata}},\ }\href@noop {} {\bibfield  {journal} {\bibinfo  {journal} {J.
  Phys. Soc. Jpn.}\ }\textbf {\bibinfo {volume} {66}},\ \bibinfo {pages} {2221}
  (\bibinfo {year} {1997})}\BibitemShut {NoStop}%
\bibitem [{\citenamefont {Ruiz}\ \emph {et~al.}(1997)\citenamefont {Ruiz},
  \citenamefont {Alemany}, \citenamefont {Alvarez},\ and\ \citenamefont
  {Cano}}]{ruiz97_2}%
  \BibitemOpen
  \bibfield  {author} {\bibinfo {author} {\bibfnamefont {E.}~\bibnamefont
  {Ruiz}}, \bibinfo {author} {\bibfnamefont {P.}~\bibnamefont {Alemany}},
  \bibinfo {author} {\bibfnamefont {S.}~\bibnamefont {Alvarez}}, \ and\
  \bibinfo {author} {\bibfnamefont {J.}~\bibnamefont {Cano}},\ }\href@noop {}
  {\bibfield  {journal} {\bibinfo  {journal} {Inorg. Chem.}\ }\textbf {\bibinfo
  {volume} {36}},\ \bibinfo {pages} {3683} (\bibinfo {year}
  {1997})}\BibitemShut {NoStop}%
\bibitem [{\citenamefont {Okubo}\ \emph {et~al.}(2009)\citenamefont {Okubo},
  \citenamefont {Yamamoto}, \citenamefont {Fujisawa}, \citenamefont {Ohta},
  \citenamefont {Nakamura},\ and\ \citenamefont {Kikuchi}}]{antlerite_2009}%
  \BibitemOpen
  \bibfield  {author} {\bibinfo {author} {\bibfnamefont {S.}~\bibnamefont
  {Okubo}}, \bibinfo {author} {\bibfnamefont {H.}~\bibnamefont {Yamamoto}},
  \bibinfo {author} {\bibfnamefont {M.}~\bibnamefont {Fujisawa}}, \bibinfo
  {author} {\bibfnamefont {H.}~\bibnamefont {Ohta}}, \bibinfo {author}
  {\bibfnamefont {T.}~\bibnamefont {Nakamura}}, \ and\ \bibinfo {author}
  {\bibfnamefont {H.}~\bibnamefont {Kikuchi}},\ }\href
  {http://stacks.iop.org/1742-6596/150/i=4/a=042156} {\bibfield  {journal}
  {\bibinfo  {journal} {J. Phys.: Conference Series}\ }\textbf {\bibinfo
  {volume} {150}},\ \bibinfo {pages} {042156} (\bibinfo {year}
  {2009})}\BibitemShut {NoStop}%
\bibitem [{\citenamefont {Hara}\ \emph {et~al.}(2011)\citenamefont {Hara},
  \citenamefont {Kondo},\ and\ \citenamefont {Sato}}]{antlerite_2011}%
  \BibitemOpen
  \bibfield  {author} {\bibinfo {author} {\bibfnamefont {S.}~\bibnamefont
  {Hara}}, \bibinfo {author} {\bibfnamefont {H.}~\bibnamefont {Kondo}}, \ and\
  \bibinfo {author} {\bibfnamefont {H.}~\bibnamefont {Sato}},\ }\href {\doibase
  10.1143/JPSJ.80.043701} {\bibfield  {journal} {\bibinfo  {journal} {Journal
  of the Physical Society of Japan}\ }\textbf {\bibinfo {volume} {80}},\
  \bibinfo {pages} {043701} (\bibinfo {year} {2011})}\BibitemShut {NoStop}%
\bibitem [{\citenamefont {Koo}\ \emph {et~al.}(2012)\citenamefont {Koo},
  \citenamefont {Kremer},\ and\ \citenamefont {Whangbo}}]{antlerite_2012}%
  \BibitemOpen
  \bibfield  {author} {\bibinfo {author} {\bibfnamefont {H.-J.}\ \bibnamefont
  {Koo}}, \bibinfo {author} {\bibfnamefont {R.~K.}\ \bibnamefont {Kremer}}, \
  and\ \bibinfo {author} {\bibfnamefont {M.-H.}\ \bibnamefont {Whangbo}},\
  }\href {\doibase 10.1143/JPSJ.81.063704} {\bibfield  {journal} {\bibinfo
  {journal} {J. Phys. Soc. Jpn.}\ }\textbf {\bibinfo {volume} {81}},\ \bibinfo
  {pages} {063704} (\bibinfo {year} {2012})}\BibitemShut {NoStop}%
\bibitem [{\citenamefont {Fujii}\ \emph {et~al.}(2013)\citenamefont {Fujii},
  \citenamefont {Ishikawa}, \citenamefont {Kikuchi}, \citenamefont {Narumi},
  \citenamefont {Nojiri}, \citenamefont {Hara},\ and\ \citenamefont
  {Sato}}]{antlerite_2013}%
  \BibitemOpen
  \bibfield  {author} {\bibinfo {author} {\bibfnamefont {Y.}~\bibnamefont
  {Fujii}}, \bibinfo {author} {\bibfnamefont {Y.}~\bibnamefont {Ishikawa}},
  \bibinfo {author} {\bibfnamefont {H.}~\bibnamefont {Kikuchi}}, \bibinfo
  {author} {\bibfnamefont {Y.}~\bibnamefont {Narumi}}, \bibinfo {author}
  {\bibfnamefont {H.}~\bibnamefont {Nojiri}}, \bibinfo {author} {\bibfnamefont
  {S.}~\bibnamefont {Hara}}, \ and\ \bibinfo {author} {\bibfnamefont
  {H.}~\bibnamefont {Sato}},\ }\href {\doibase 10.3938/jkps.62.2054} {\bibfield
   {journal} {\bibinfo  {journal} {J. Korean Phys. Soc.}\ }\textbf {\bibinfo
  {volume} {62}},\ \bibinfo {pages} {2054} (\bibinfo {year}
  {2013})}\BibitemShut {NoStop}%
\bibitem [{\citenamefont {Goodenough}(1955)}]{gka1}%
  \BibitemOpen
  \bibfield  {author} {\bibinfo {author} {\bibfnamefont {J.~B.}\ \bibnamefont
  {Goodenough}},\ }\href {\doibase 10.1103/PhysRev.100.564} {\bibfield
  {journal} {\bibinfo  {journal} {Phys. Rev.}\ }\textbf {\bibinfo {volume}
  {100}},\ \bibinfo {pages} {564} (\bibinfo {year} {1955})}\BibitemShut
  {NoStop}%
\bibitem [{\citenamefont {Kanamori}(1959)}]{gka2}%
  \BibitemOpen
  \bibfield  {author} {\bibinfo {author} {\bibfnamefont {J.}~\bibnamefont
  {Kanamori}},\ }\href {\doibase 10.1016/0022-3697(59)90061-7} {\bibfield
  {journal} {\bibinfo  {journal} {J. Phys. Chem. Solids}\ }\textbf {\bibinfo
  {volume} {10}},\ \bibinfo {pages} {87} (\bibinfo {year} {1959})}\BibitemShut
  {NoStop}%
\bibitem [{\citenamefont {Anderson}(1963)}]{gka3}%
  \BibitemOpen
  \bibfield  {author} {\bibinfo {author} {\bibfnamefont {P.~W.}\ \bibnamefont
  {Anderson}},\ }\href@noop {} {\bibfield  {journal} {\bibinfo  {journal}
  {Solid State Phys.}\ }\textbf {\bibinfo {volume} {14}},\ \bibinfo {pages}
  {99} (\bibinfo {year} {1963})}\BibitemShut {NoStop}%
\bibitem [{\citenamefont {Janson}\ \emph {et~al.}(2012)\citenamefont {Janson},
  \citenamefont {Rousochatzakis}, \citenamefont {Tsirlin}, \citenamefont
  {Richter}, \citenamefont {Skourski},\ and\ \citenamefont
  {Rosner}}]{decoratedSS}%
  \BibitemOpen
  \bibfield  {author} {\bibinfo {author} {\bibfnamefont {O.}~\bibnamefont
  {Janson}}, \bibinfo {author} {\bibfnamefont {I.}~\bibnamefont
  {Rousochatzakis}}, \bibinfo {author} {\bibfnamefont {A.~A.}\ \bibnamefont
  {Tsirlin}}, \bibinfo {author} {\bibfnamefont {J.}~\bibnamefont {Richter}},
  \bibinfo {author} {\bibfnamefont {Y.}~\bibnamefont {Skourski}}, \ and\
  \bibinfo {author} {\bibfnamefont {H.}~\bibnamefont {Rosner}},\ }\href
  {\doibase 10.1103/PhysRevB.85.064404} {\bibfield  {journal} {\bibinfo
  {journal} {Phys. Rev. B}\ }\textbf {\bibinfo {volume} {85}},\ \bibinfo
  {pages} {064404} (\bibinfo {year} {2012})}\BibitemShut {NoStop}%
\bibitem [{\citenamefont {Maeshima}\ and\ \citenamefont
  {Okunishi}(2000)}]{j1j2_AFM_dmrg}%
  \BibitemOpen
  \bibfield  {author} {\bibinfo {author} {\bibfnamefont {N.}~\bibnamefont
  {Maeshima}}\ and\ \bibinfo {author} {\bibfnamefont {K.}~\bibnamefont
  {Okunishi}},\ }\href {\doibase 10.1103/PhysRevB.62.934} {\bibfield  {journal}
  {\bibinfo  {journal} {Phys. Rev. B}\ }\textbf {\bibinfo {volume} {62}},\
  \bibinfo {pages} {934} (\bibinfo {year} {2000})}\BibitemShut {NoStop}%
\bibitem [{\citenamefont {Johnston}\ \emph {et~al.}(2000)\citenamefont
  {Johnston}, \citenamefont {Kremer}, \citenamefont {Troyer}, \citenamefont
  {Wang}, \citenamefont {Kl\"umper}, \citenamefont {Bud'ko}, \citenamefont
  {Panchula},\ and\ \citenamefont {Canfield}}]{johnston2000}%
  \BibitemOpen
  \bibfield  {author} {\bibinfo {author} {\bibfnamefont {D.~C.}\ \bibnamefont
  {Johnston}}, \bibinfo {author} {\bibfnamefont {R.~K.}\ \bibnamefont
  {Kremer}}, \bibinfo {author} {\bibfnamefont {M.}~\bibnamefont {Troyer}},
  \bibinfo {author} {\bibfnamefont {X.}~\bibnamefont {Wang}}, \bibinfo {author}
  {\bibfnamefont {A.}~\bibnamefont {Kl\"umper}}, \bibinfo {author}
  {\bibfnamefont {S.~L.}\ \bibnamefont {Bud'ko}}, \bibinfo {author}
  {\bibfnamefont {A.~F.}\ \bibnamefont {Panchula}}, \ and\ \bibinfo {author}
  {\bibfnamefont {P.~C.}\ \bibnamefont {Canfield}},\ }\href {\doibase
  10.1103/PhysRevB.61.9558} {\bibfield  {journal} {\bibinfo  {journal} {Phys.
  Rev. B}\ }\textbf {\bibinfo {volume} {61}},\ \bibinfo {pages} {9558}
  (\bibinfo {year} {2000})}\BibitemShut {NoStop}%
\bibitem [{\citenamefont {Lebernegg}\ \emph {et~al.}(2016)\citenamefont
  {Lebernegg}, \citenamefont {Tsirlin}, \citenamefont {Janson}, \citenamefont
  {Redhammer},\ and\ \citenamefont {Rosner}}]{langite}%
  \BibitemOpen
  \bibfield  {author} {\bibinfo {author} {\bibfnamefont {S.}~\bibnamefont
  {Lebernegg}}, \bibinfo {author} {\bibfnamefont {A.~A.}\ \bibnamefont
  {Tsirlin}}, \bibinfo {author} {\bibfnamefont {O.}~\bibnamefont {Janson}},
  \bibinfo {author} {\bibfnamefont {G.~J.}\ \bibnamefont {Redhammer}}, \ and\
  \bibinfo {author} {\bibfnamefont {H.}~\bibnamefont {Rosner}},\ }\href@noop {}
  {\bibfield  {journal} {\bibinfo  {journal} {New J. Phys.}\ }\textbf {\bibinfo
  {volume} {18}},\ \bibinfo {pages} {033020} (\bibinfo {year}
  {2016})}\BibitemShut {NoStop}%
\end{thebibliography}
%

\clearpage

\begin{table*} [h!]
\begin{tabular}{c}
\huge{\texttt{Supporting Material}} \\
\end{tabular}
\end{table*}

\begin{widetext}

\clearpage

\begin{figure} [h]
\includegraphics[width=13cm]{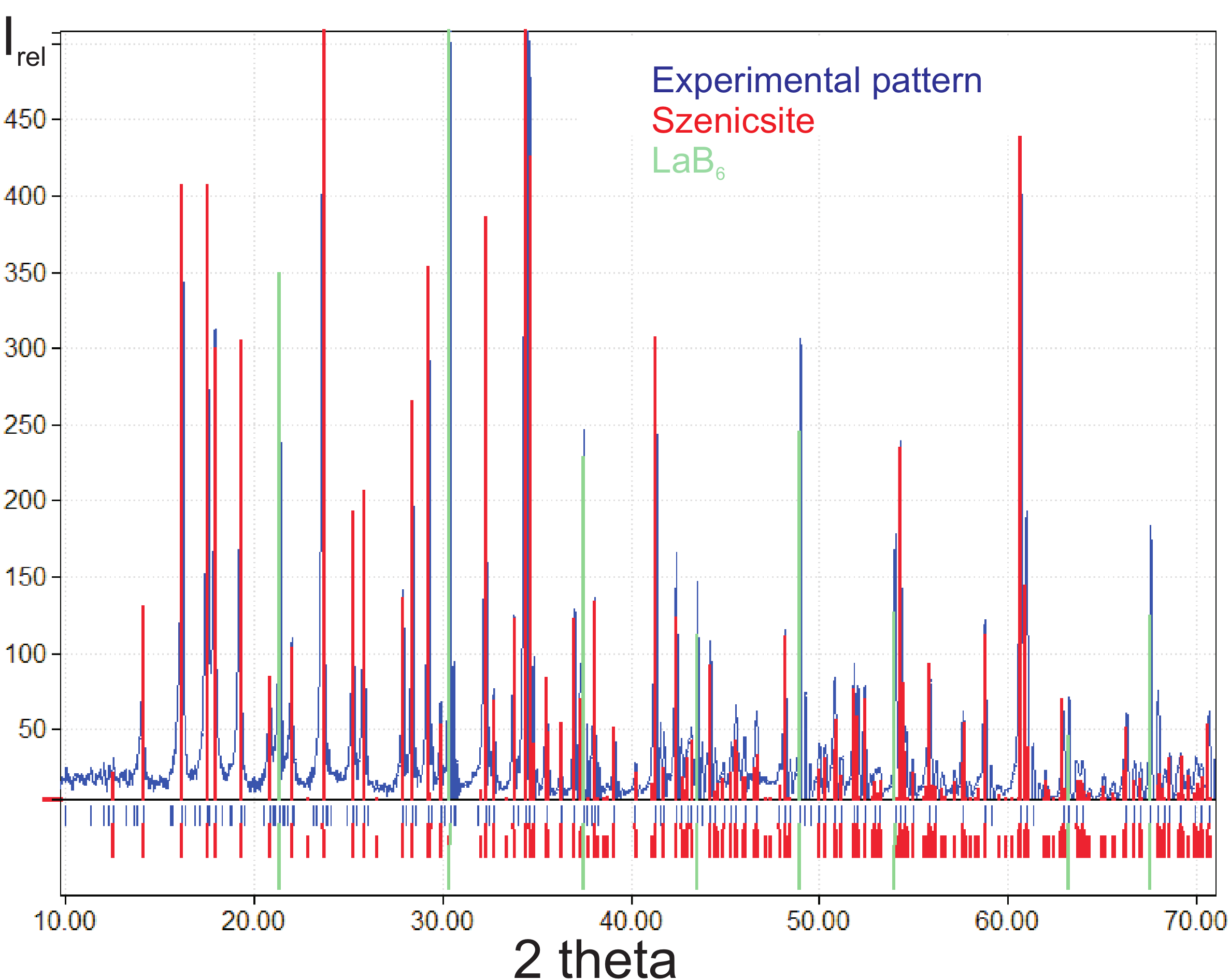}
\caption{\label{xrd}
(Color online) Room temperature powder X-ray diffraction pattern (Huber G670 Guinier camera, CuK$_{\alpha\,1}$ radiation, ImagePlate detector, $2\theta\,=\,3-100^{\circ}$ angle range) of the szenicsite sample. LaB$_6$ was used as an internal standard. No other phases are detectable.}
\end{figure}

\begin{figure} [h]
\includegraphics{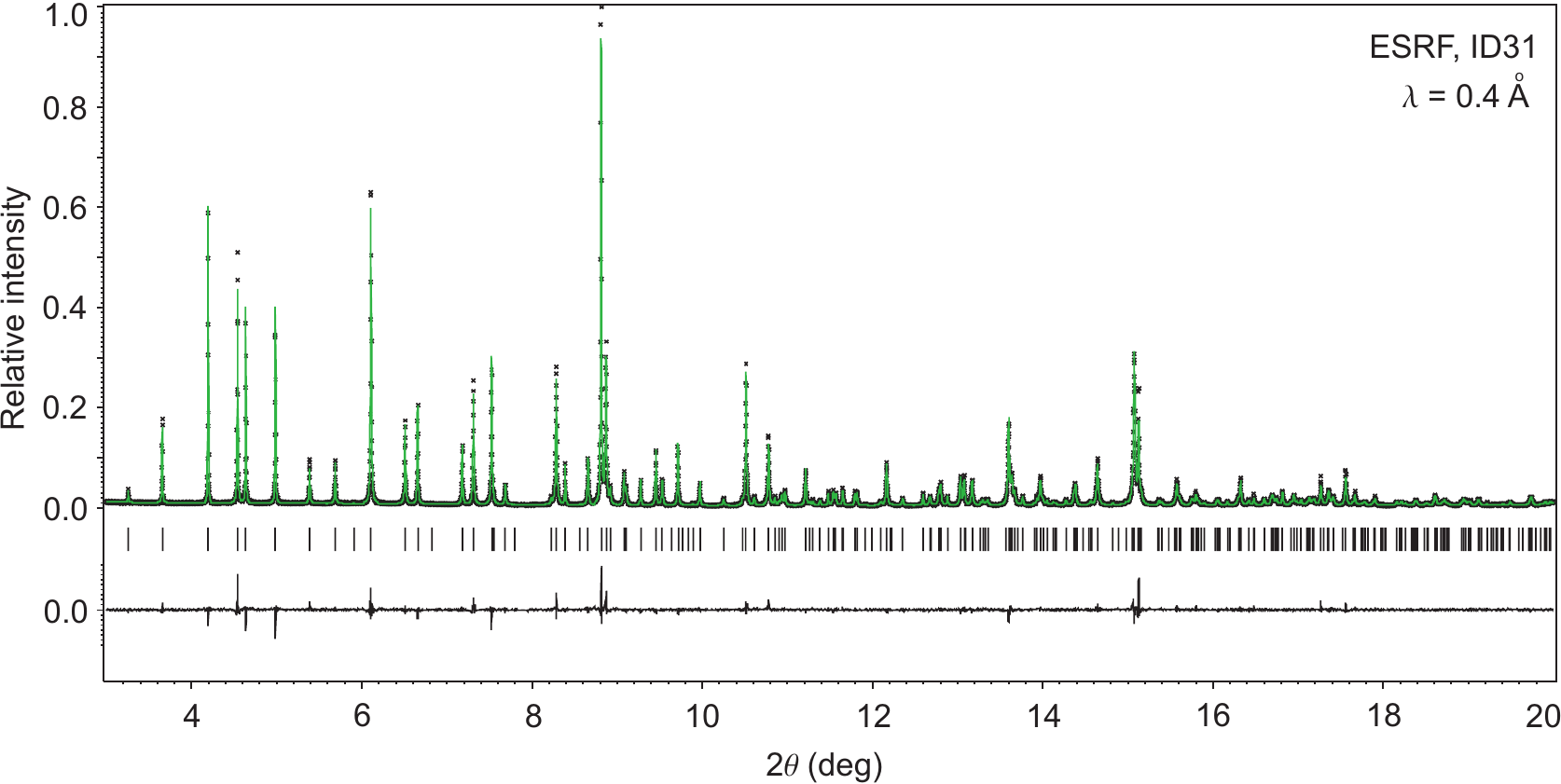}
\caption{
Rietveld refinement for szenicsite of the room temperature high-resolution XRD data. Ticks show the reflection positions of szenicsite.}
\end{figure}

\begin{figure} [h]
\includegraphics{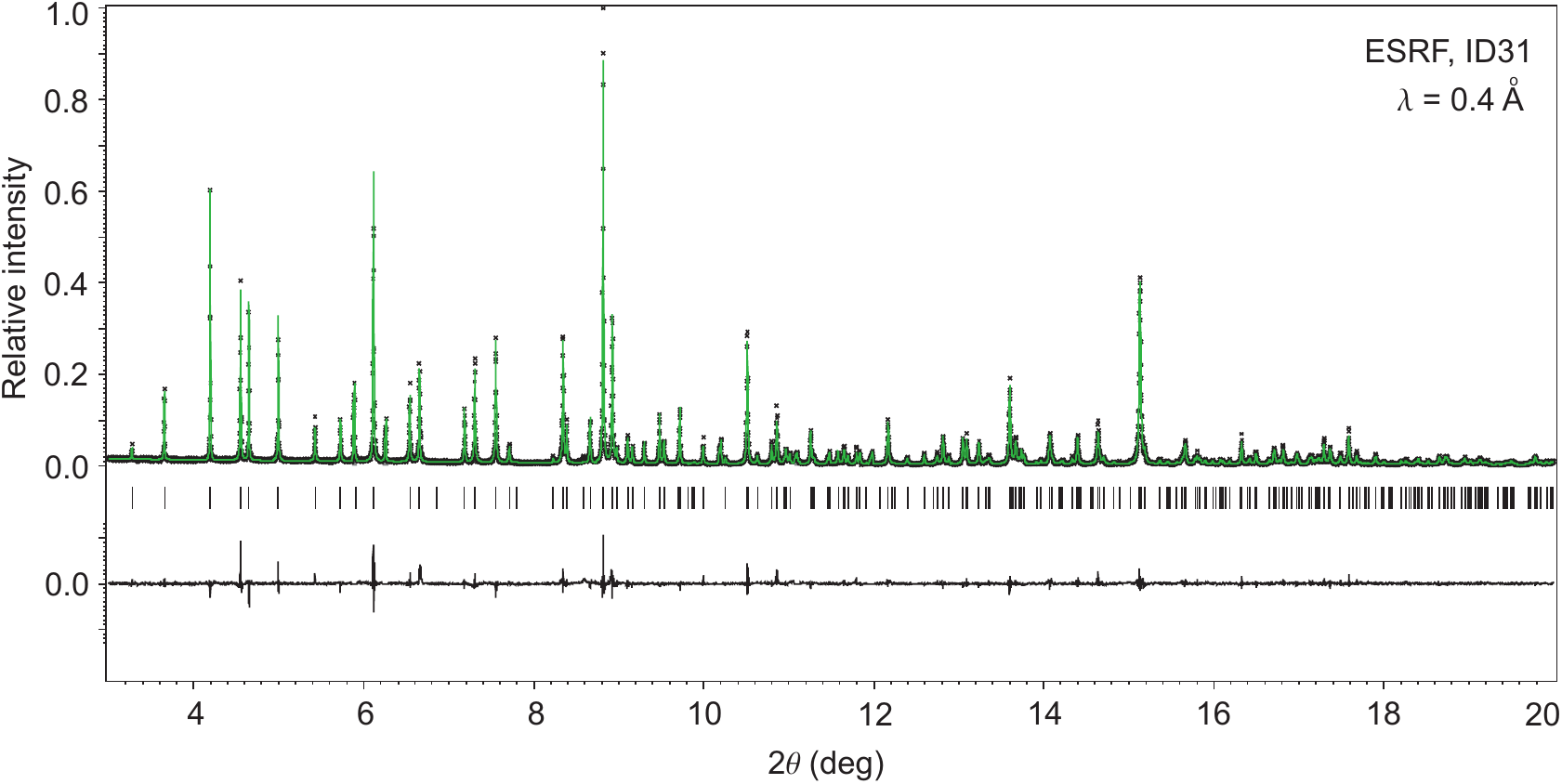}
\caption{
Rietveld refinement for szenicsite of the 80\,K high-resolution XRD data. Ticks show the reflection positions of szenicsite.}
\end{figure}

\begin{table}
\caption{
Atomic positions in szenicsite refined from synchrotron powder diffraction data collected at 80\,K (upper lines) and at room temperature (lower lines). Atomic displacement parameters $U_{\rm iso}$ are given in units of $10^{-2}$\,\r A$^{-2}$. The hydrogen positions are not listed, because they were not refined (see also Table~I in the main text). Atomic displacement parameters of oxygen were refined as a single parameter. The lattice parameters are $a=8.45745(7)$\,\r A, $b=12.5622(1)$\,\r A, and $c=6.07980(4)$\,\r A at 80\,K ($R_I=0.044$, $R_p=0.126$) and $a=8.52196(4)$\,\r A, $b=12.5581(1)$\,\r A, and $c=6.07954(2)$\,\r A ($R_I=0.0433$, $R_p=0.088$) at room temperature.
}
\begin{minipage}{12cm}
\begin{ruledtabular}
\begin{tabular}{cccccc}
    &      & $x/a$      & $y/b$      & $z/c$        & $U_{\rm iso}$ \\\hline
Cu1 & $8h$ & 0.2678(1)  & 0.1330(1)  & $-0.2506(3)$ & 0.24(3) \\
    &      & 0.26883(9) & 0.13287(7) & $-0.2504(2)$ & 0.74(2) \\
Cu2 & $2b$ & 0          & 0          & 0.5          & 0.43(7) \\
    &      & 0          & 0          & 0.5          & 0.81(5) \\
Cu3 & $2a$ & 0          & 0          & 0            & 0.41(8) \\
    &      & 0          & 0          & 0            & 1.04(5) \\
Mo  & $4g$ & 0.1242(1)  & 0.3688(1)  & 0            & 0.47(3) \\
    &      & 0.1245(1)  & 0.36939(8) & 0            & 0.97(2) \\
O1  & $4g$ & 0.2670(12) & 0.2728(7)  & 0            & 0.07(7) \\
    &      & 0.2671(8)  & 0.2671(5)  & 0            & 0.98(5) \\
O2  & $8h$ & 0.0092(6)  & 0.3583(5)  & 0.2442(11)   & 0.07(7) \\
    &      & 0.0111(4)  & 0.3581(3)  & 0.2450(7)    & 0.98(5) \\
O3  & $4g$ & 0.2262(12) & 0.4948(8)  & 0            & 0.07(7) \\
    &      & 0.2230(8)  & 0.4952(5)  & 0            & 0.98(5) \\
O4  & $4g$ & 0.2760(12) & 0.0347(8)  & 0            & 0.07(7) \\
    &      & 0.2726(8)  & 0.0388(5)  & 0            & 0.98(5) \\
O5  & $8g$ & 0.0290(7)  & 0.1004(4)  & $-0.2508(12)$& 0.07(7) \\
    &      & 0.0274(4)  & 0.1001(3)  & $-0.2483(7)$ & 0.98(5) \\
O6  & $4g$ & 0.2615(12) & 0.2213(8)  & 0.5          & 0.07(7) \\
    &      & 0.2613(8)  & 0.2226(5)  & 0.5          & 0.98(5) \\
\end{tabular}
\end{ruledtabular}
\end{minipage}
\end{table}

\begin{figure}[h]
\includegraphics[width=12cm]{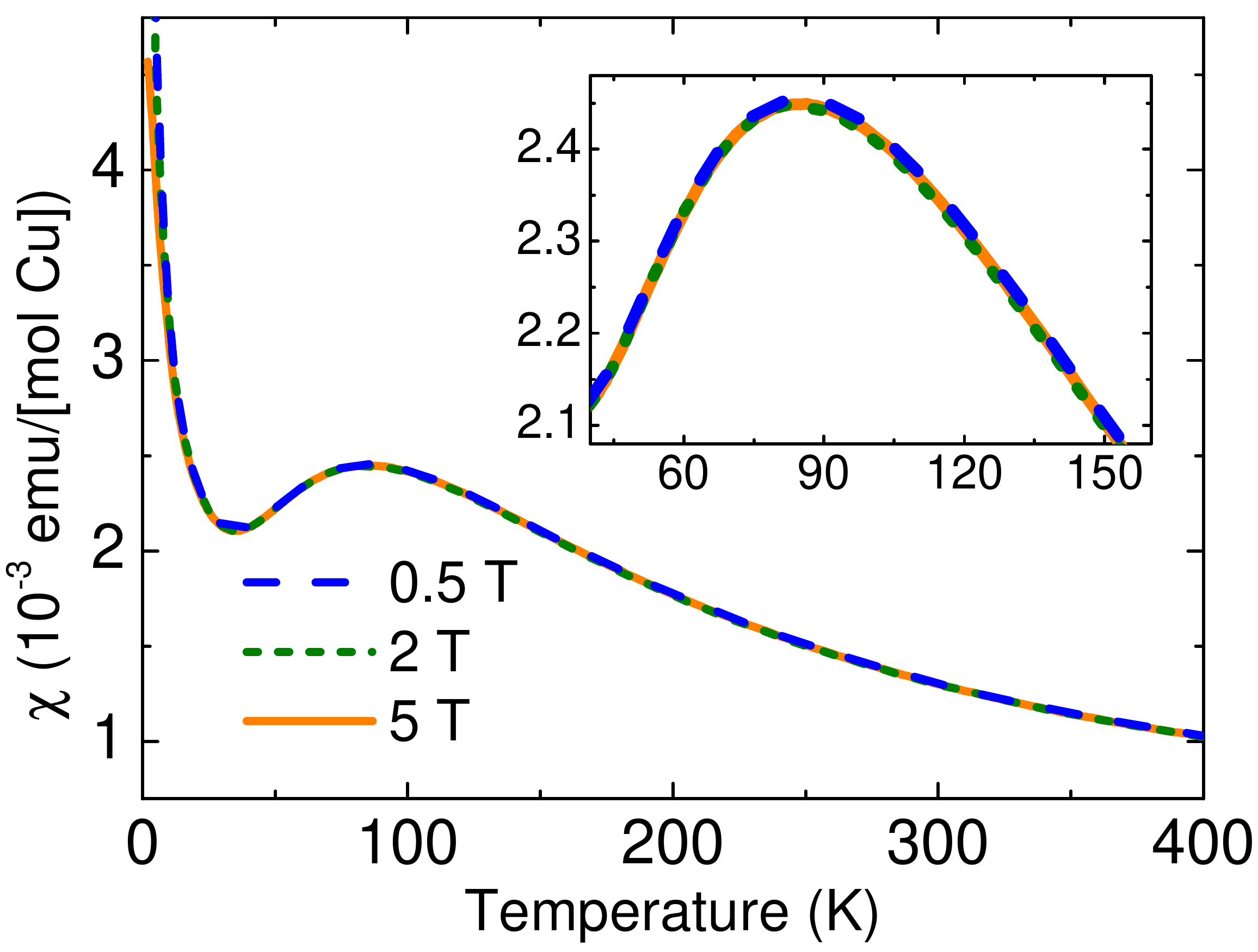}
\caption{\label{sus_H}
(Color online) The magnetic susceptibility $\chi(T)$ of szenicsite measured in three different magnetic fields. Only the low-temperature part is slightly field-dependent, which arises from paramagnetic imperfections, while the region of the maximum is unaffected.}
\end{figure}

\begin{figure}[h]
\includegraphics[width=12cm]{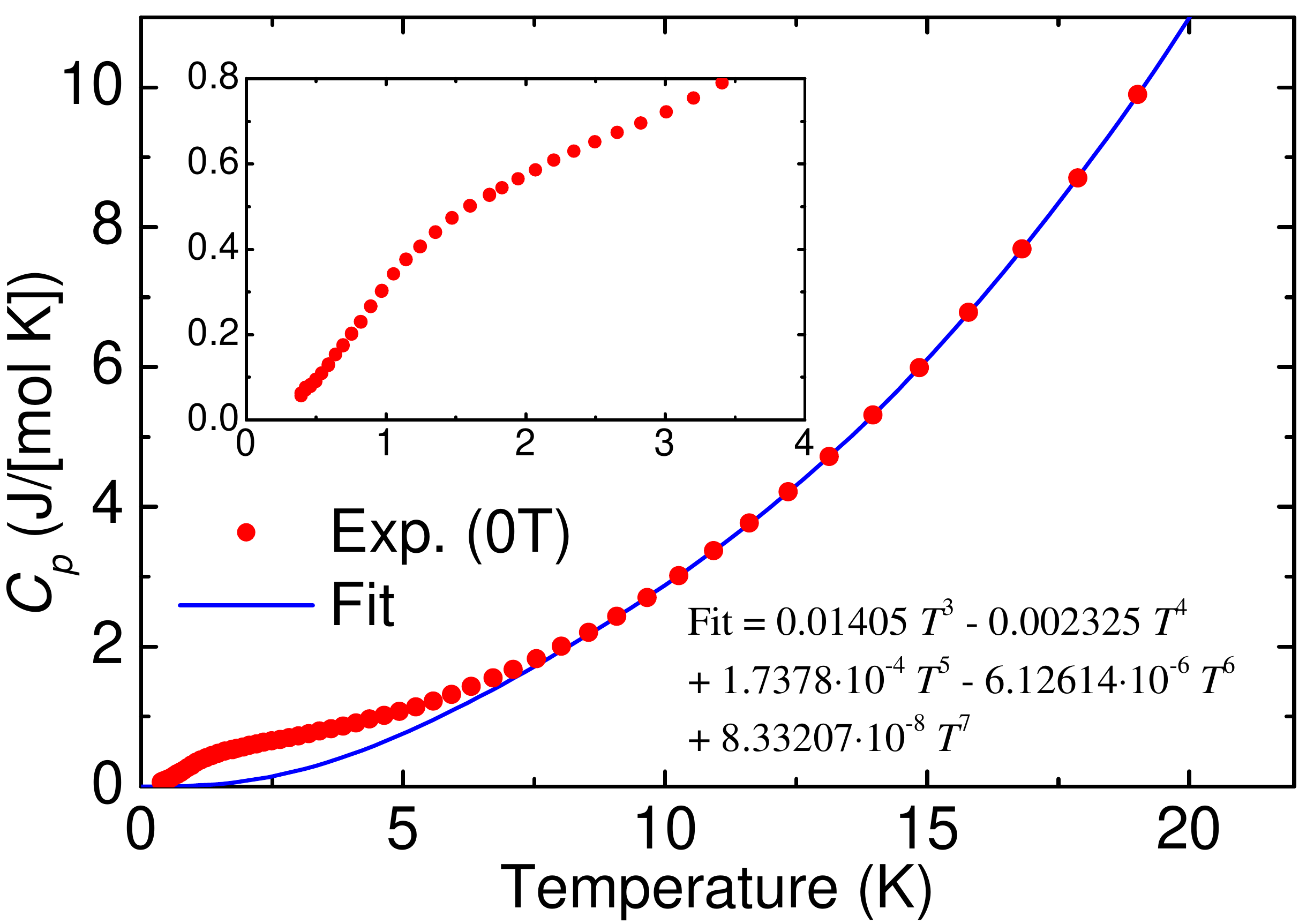}
\caption{\label{sus_H2}
(Color online) The specific heat data $C_p(T)$ of szenicsite measured between 0.5--20\,K in a magnetic field of 0\,T. The blue line shows the lattice background $C_{lat}(T)$ which we obtained by fitting a polynomial of the form $C_{lat}(T)=\sum \limits_{n=3}^{n=7} c_nT^n$ to the experimental data in the temperature regime 15--39\,K. $C_p(T)-C_{lat}(T)$ yields the magnetic contribution $C_{mag}$ to the specific heat. The inset is a blow-up of the low-temperature $C_p(T)$ data.}
\end{figure}

\end{widetext}

\end{document}